\documentclass{aa}
\usepackage{graphicx}	% Including figure files

\usepackage{amsmath}	% Advanced maths commands
\usepackage{amssymb}	% Extra maths symbols
\usepackage{multicol}   % Multi-column entries in tables
\usepackage{color}
\usepackage{xspace}
\usepackage{subfigure}
\usepackage{CJKutf8}
\usepackage{txfonts}
\usepackage{natbib,twoopt}
\usepackage[colorlinks=true,
            linkcolor=blue,
            urlcolor=blue,
            citecolor=blue,
            anchorcolor=blue]{hyperref}
\usepackage{ulem}
\makeatletter
\renewcommand*\aa@pageof{, page \thepage{} of \pageref*{LastPage}}
\makeatother
\usepackage{ulem,bm}
\definecolor{mygray}{gray}{0.6}
\definecolor{TsinghuaPurple}{cmyk}{0.58,0.90,0,0}
\definecolor{magenta}{rgb}{0.858, 0.188, 0.478}

\makeatletter
\def\uwave{\bgroup \markoverwith{\lower3.5\p@\hbox{\sixly \textcolor{red}{\char58}}}\ULon}
\font\sixly=lasy6 % does not re-load if already loaded, so no memory problem.
\makeatother

\definecolor{emerald}{RGB}{0,155,155}
\newcommand{\hjc}[1]{\textcolor{emerald}{[\small HJ: \textit{\small #1}]}}
\newcommand{\hjadd}[1]{\textcolor{emerald}{#1}}
\newcommand{\hjch}[2]{\textcolor{emerald}{#2}}
\newcommand{\hjrem}[1]{\textcolor{mygray}{\sout{#1}}}

\renewcommand{\hjc}[1]{}
\renewcommand{\hjadd}[1]{#1}
\renewcommand{\hjch}[2]{{#2}}
\renewcommand{\hjrem}[1]{}
\newcommand{\xxx}[1]{\textcolor{blue}{\textbf{xxx}\xspace}}
\newcommand{\fg}[1]{Fig.~\ref{fig:#1}}
\newcommand{\Fg}[1]{Figure~\ref{fig:#1}}%beginning of the sentence

\newcommand{\eq}[1]{Eq.~(\ref{eq:#1})\xspace}
\newcommand{\Eq}[1]{Equation~(\ref{eq:#1})\xspace}%beginning of the sentence
\newcommand{\eqs}[2]{Eqs.\ (\ref{eq:#1}) and (\ref{eq:#2})}

\newcommand{\tb}[1]{Table~\ref{tab:#1}\xspace}
\newcommand{\Tb}[1]{Table~\ref{tab:#1}\xspace}%beginning of the sentence
\newcommand{\se}[1]{Sect.~\ref{sec:#1}\xspace}
\newcommand{\Se}[1]{Section~\ref{sec:#1}\xspace}%beginning of the sentence

\newcommand{\App}[1]{Appendix~\ref{app:#1}\xspace}

\begin{document}

    \title{Chemical Footprints of Giant Planet Formation.}
    \subtitle{Role of Planet Accretion in Shaping the C/O Ratio of Protoplanetary Disks}
    \authorrunning{H. Jiang, Y. Wang, C.W. Ormel, S. Krijt, R. Dong}

    \author{
    Haochang Jiang (\begin{CJK*}{UTF8}{gbsn}蒋昊昌\end{CJK*})
    \inst{1}\fnmsep\inst{2},
    Yu Wang (\begin{CJK*}{UTF8}{gbsn}王雨\end{CJK*})\inst{2},
    Chris W. Ormel\inst{2},
    Sebastiaan Krijt\inst{3},
    Ruobing Dong (\begin{CJK*}{UTF8}{gbsn}董若冰\end{CJK*})\inst{4}
    }
    
    \institute{
    European Southern Observatory,
    Karl-Schwarzschild-Str 2, 85748 Garching, Germany
    \and
    Department of Astronomy, Tsinghua University, 30 Shuangqing Rd,
    Haidian DS 100084, Beijing, China
    \and
    School of Physics and Astronomy, University of Exeter,
    Stocker Road, Exeter EX4 4QL, UK
    \and
    Department of Physics and Astronomy, University of Victoria, Victoria, BC V8P 5C2, Canada\\
    \email{hjiang@eso.org; wang-y21@mails.tsinghua.edu.cn}
    }

   \date{Received February 1, 1997; accepted March 23, 1997}

% \abstract{}{}{}{}{} 
% 5 {} token are mandatory
 
  \abstract
  % context heading (optional)
  % {} leave it empty if necessary  
   {Protoplanetary disks, the birthplaces of planets, commonly feature bright rings and dark gaps in both continuum and line emission maps. Accreting planets are interacting with the disk, not only through gravity, but also by changing the local irradiation and elemental abundances, which are essential ingredients for disk chemistry.}
  % aims heading (mandatory)
   {We propose that giant planet accretion can leave chemical footprints in the gas local to the planet, which potentially leads to the spatial coincidence of molecular emissions with the planet in ALMA observation.}
  % methods heading (mandatory)
   {Through 2D multi-fluid hydrodynamical simulations in Athena++ with built-in sublimation, we simulate the process of an accreting planet locally heating up its vicinity, opening a gas gap in the disk, and creating the conditions for C-photochemistry. }
  % results heading (mandatory)
   {An accreting planet located outside the methane snowline can render the surrounding gas hot enough to sublimate the C-rich organics off pebbles before they are accreted by the planet. This locally elevates the disk gas-phase C/O ratio, providing a potential explanation for the C$_2$H line-emission rings observed with ALMA. In particular, our findings provide an explanation for the MWC\,480 disk, where previous work has identified a statistically significant spatial coincidence of line-emission rings inside a continuum gap. 
   }
  % conclusions heading (optional), leave it empty if necessary 
   {Our findings present a novel view of linking the gas accretion of giant planets and their natal disks through the chemistry signals. This model demonstrates that giant planets can actively shape their forming chemical environment, moving beyond the traditional understanding of the direct mapping of primordial disk chemistry onto planets.}


   \maketitle
%
%-------------------------------------------------------------------

\section{Introduction}

Protoplanetary disks are the cradle of planets. Imaging observations with high resolution and high contrast using instruments such as ALMA and Gemini/GPI have found that most large disks ($r>50$\,au), if not all, exhibit substructures such as bright rings and dark gaps \citep[e.g.,][]{AndrewsEtal2018b,LongEtal2018f,vanderMarelEtal2019}. It is widely speculated that planet-disk interactions are responsible for many of these substructures \citep[e.g, ][]{ZhuEtal2011,DongEtal2015,RosottiEtal2016,ZhangEtal2018,PaardekooperEtal2022}. Nevertheless, detecting planets in these young systems is challenging \citep[][]{XieEtal2020,HuelamoEtal2022,FolletteEtal2022}. Besides PDS\,70 \citep{KepplerEtal2018,HaffertEtal2019} and AB\,Aur \citep{CurrieEtal2022,ZhouEtal2022}, most claims of planet detections are indirect \citep[e.g.,][]{PinteEtal2019,PinteEtal2020,SpeedieEtal2022b,IzquierdoEtal2022} and might be debatable \citep[][]{SpeedieDong2022,FedeleEtal2023}.

Among the indirect planet detection method, the role of astrochemistry is gaining increasing prominence with the development of facilities such as ALMA and JWST \citep[e.g.,][]{OebergBergin2021}. The classical core accretion planet formation dictates that a Neptune-mass planet will quickly grow into a Jovian-mass gas giant during a runaway growth phase, in which gas freely falls into its gravitational potential. During the runaway phase the planet's mass and accretion rate $\dot{M}_{\rm p}$ exponentially grow, making the protoplanet an irradiating source in the disk \citep[e.g.,][]{Rafikov2006,MordasiniEtal2012}.

Luminous young planets locally heat up their vicinity, resulting in efficient thermal sublimation of molecular species that are originally frozen out. 
Therefore, 
localized submillimeter molecular emission might appear around the planet as proposed in \citet{CleevesEtal2015}. This scenario is particularly compelling for the recently identified local emission enhancements of SO in HD\,100546 \citep{BoothEtal2023}, and $^{13}$CO in AS\,209 \citep{BaeEtal2022}. 
Yet, the calculation of \citet{CleevesEtal2015} used a static gas density profile, where the differential rotation of material around the planet is ignored. For a gap-opening planet at distant $r_0$, as the planet opens a wider gap with width $\Delta w$ as it is accreting, the synodic timescale around the planet gap $r_0/\Delta w\Omega_K$, where $\Omega_K$ is the Keplerian frequency, gets shorter. If this timescale is shorter than the typical chemical timescale, materials can be sheared around the co-orbital region of the accreting planet, and the asymmetric planet-heated chemical emission might be elongated and become arc- or even ring-like.

Thanks to the high spatial resolution and high sensitivity of ALMA, ring-like substructures are found to be ubiquitous in line-emission maps as well \citep[e.g.,][]{OebergEtal2021}. Among those species with line-emission detection, the radial C$_2$H, one of the most well-studied hydrocarbons, is believed an indicate an active photo-chemistry, driven by high UV-fluxes and gas with a C/O ratio exceeding unity \citep{BerginEtal2016,KamaEtal2016,MiotelloEtal2019,BergnerEtal2019,AlarconEtal2021,vanderMarelEtal2021c,BosmanEtal2021b}. C$_2$H can serve as a precursor for larger molecule formation through chemical reaction network \citep[e.g.,][]{BastEtal2013,OebergBergin2021}. Alternatively, C$_2$H may also be a product of top-down photo-dissociation hydrocarbon chemistry \citep{GuzmanEtal2015}.

Due to the specific formation conditions required for C$_2$H --high C/O ratio and high UV penetration-- and its bridging role in the chemical reaction network, we aim to explore whether a young giant planet can be responsible for achieving these conditions, and discuss how the chemical footprint of the accretion of the planet will manifest in protoplanetary disks. 

In this study, we conduct 2D multi-fluid hydrodynamical simulations that include built-in sublimation to simulate an accreting planet in Athena++ \citep{HuangBai2022,WangEtal2023}. The planet is modeled as a sink particle in the simulation with self-consistent accretion prescription. As the planet accretes, the potential energy of the accreted gas is released as the planet's luminosity, locally heating the gas surrounding the planet and releasing C-rich volatile, whose abundance we follow in Athena++. We then post-process our simulation outputs with an empirical formula derived from chemical network simulations on C$_2$H formation. In addition, we compare our simulation results with line-emission observations from ALMA.

The plan of the paper is as follows. The theoretical background and numerical setup are described in \Se{model}. We present and explain the numerical results in \Se{results}. The results are further discussed in \Se{discussion}. We summarize the main results and conclusions in \Se{conclusions}.

%--------------------------------------------------------------------
\section{Model}\label{sec:model}

We simulate the planet and disk interaction with multi-fluid 2D hydrodynamics simulations in polar coordinates (radius $r$ and azimuth $\phi$). 
The disk consists of three mass components -- non-condensable gas, ice-rich pebbles, and sublimated vapor. To model them, we introduce gas fluid (``g''), particle fluid (``p'') and tracer fluid (``tr'') respectively. The “gas fluid” is the mixture of non-condensable gaseous species (mostly H and He) and sublimated vapor. There is only one gas fluid. In addition, the density of each vapor species is followed by a “tracer fluid”, which inherits the hydrodynamical properties (i.e., velocity field) of the gas fluid. Pebbles are modelled as compound of multiple ``particle fluids'', each of which represents a solid component (i.e., ice and refractory). With the recently developed multi-dust fluid module of \texttt{Athena++} \citep{HuangBai2022}, the gas and ice can co-evolve in a self-consistent manner. 
The governing equations read (see also Sec.~2.2 of \cite{WangEtal2023}),

\begin{equation}\label{eq:CE_gas}
    \frac{\partial \rho_{\mathrm{g}}}{\partial t} + \nabla \cdot (\rho_{\mathrm{g}} \bm{v}_{\mathrm{g}}) = 0,
\end{equation}

\begin{equation}
\label{eq:ME_gas}
    \frac{\partial (\rho_{\mathrm{g}} \bm{v}_{\mathrm{g}})}{\partial t} + \nabla \cdot (\rho_{\mathrm{g}} \bm{v}_{\mathrm{g}} \bm{v}_{\mathrm{g}} + P_{\mathrm{g}}\bm{I} + \bm{\Pi}_{\mathrm{\nu}}) = \rho_{\mathrm{g}} \bm{f}_{\mathrm{g, src}},
\end{equation}

\begin{equation}
\label{eq:CE_dust}
    \frac{\partial \rho_{\mathrm{p}}}{\partial t}+\nabla \cdot\left(\rho_{\mathrm{p}} \bm{v}_{\mathrm{p}}+ \mathcal{F}_{\mathrm{p, dif}}\right)=0,
\end{equation}

\begin{equation}
\label{eq:ME_dust}
    \begin{aligned}
    \frac{\partial \rho_{\mathrm{p}}\left(\bm{v}_{\mathrm{p}}+\bm{v}_{\mathrm{p}, \mathrm{dif}}\right)}{\partial t}+\nabla \cdot\left(\rho_{\mathrm{p}} \bm{v}_{\mathrm{p}} \bm{v}_{\mathrm{p}} + \bm{\Pi}_{\mathrm{dif}}\right) & = \\
    \rho_{\mathrm{p}} \bm{f}_{\mathrm{p}, \mathrm{src}}+\rho_{\mathrm{p}} \frac{\bm{v}_{\mathrm{g}}-\bm{v}_{\mathrm{p}}}{t_{\mathrm{s}}},
    \end{aligned}
\end{equation}

\begin{equation}
\label{eq:CE_vapor}
\frac{\partial \rho_{\mathrm{tr}}}{\partial t}+\nabla \cdot\left(\rho_{\mathrm{tr}} \bm{v}_{\mathrm{g}}+\mathcal{F}_{\mathrm{tr, dif}}\right)=0.
\end{equation}
In the above, $\rho$ is the density, $v$ is the velocity, $P_\mathrm{g}$ is the gas pressure, $\bm{I}$ is the identity tensor. The gas viscosity is described by the viscous stress tensor
\begin{equation}
    \bm{\Pi}_{\mathrm{\nu}} \equiv
    \rho_\mathrm{g}\nu_\mathrm{g}\left[\nabla\bm{v}_\mathrm{g}+\left(\nabla\bm{v}_\mathrm{g}\right)^{T}-\frac{2}{3}\left(\nabla\cdot\bm{v}_\mathrm{g}\right)\bm{I}\right]
\end{equation}
where the kinematic viscosity $\nu_\mathrm{g}$ is specifically expressed as a function of the dimensionless constant  $\alpha$ \citep{ShakuraSunyaev1973}
\begin{equation}
    \nu_\mathrm{g}=\alpha c_{s}H_{\rm g}.
\end{equation}
where  $H_{\rm g}$ is the gas scale height and $c_s$ is the local sound speed. With this, the particle concentration diffusion flux is given by
\begin{equation}\label{eq:F_p_dif}
    \mathcal{F}_{\mathrm{p, dif}} = -\rho_{\mathrm{g}}D_{\mathrm{p}} \nabla\left(\frac{\rho_\mathrm{p}}{\rho_\mathrm{g}}\right)
    \equiv \rho_{\mathrm{p}}\bm{v}_{\mathrm{p, dif}},
\end{equation}
which meanwhile defines the effective diffusive velocity $\bm{v}_{\mathrm{p, dif}}$.
And the diffusivity for dust can be prescribed as \citep{YoudinLithwick2007}
\begin{equation}
    D_{\mathrm{p}} \equiv \frac{\nu_\mathrm{g}}{1+{\rm St}^2} 
\end{equation}
with the Stokes number ${\rm St}\equiv t_\mathrm{s}\Omega_K$. $\bm{\Pi}_{\mathrm{dif}}$ is the momentum diffusion flux tensor which, combined with $\bm{v}_{\mathrm{p, dif}}$, ensures the Galilean invariance \citep{HuangBai2022} in \eq{ME_dust}. 
Similar to \eq{F_p_dif}, the tracer concentration diffusion flux is defined as,
\begin{equation}
    \mathcal{F}_{\mathrm{tr, dif}} = -\rho_{\mathrm{g}}D_{\mathrm{tr}} \nabla\left(\frac{\rho_\mathrm{p}}{\rho_\mathrm{g}}\right)
\end{equation}
with the tracer diffusivity $D_{\mathrm{tr}} = \nu_{\mathrm{g}}$, where we implicitly assume that the Schmidt number is 1. Without loss of generality, we assume the gas viscosity coefficient $\alpha$ to be $5\times10^{-4}$, and initialize the pebbles with constant Stokes number $\rm St = 0.01$ throughout the domain. \hjadd{By assuming spherical pebbles, the particles size
\begin{equation}
    s_p = \frac{2}{\pi} \frac{{\rm St}\Sigma_{\rm g}}{\rho_\bullet}
\end{equation}
is ${\sim}$0.5\,mm at the planet orbit, with internal density $\rho_\bullet = 1.67 \rm g\,cm^{-3}$ \citep{BirnstielEtal2018}.} Those values match the latest constraint approaches based on ALMA observational results \citep[see the recent review by][and references therein]{Rosotti2023}. \hjadd{In our simulations setup, the particle size $s_p$ is kept constant throughout the simulation.}
The final term in  \eq{ME_dust} represents the aerodynamic drag experienced by particles, which follow the Epstein regime throughout our simulation, with the backreaction on gas omitted in our formulation. External source terms are denoted by $f_{\mathrm{src}}$, including stellar and planetary gravity  (see \eq{grav_p}).

\subsection{Disk model}
We assume that the initial gas and pebble surface densities follow power-law distributions
\begin{equation}\label{eq:Sig}
\begin{aligned}
    \Sigma_{\rm g} &= \Sigma_{\rm g,0}\times\left(\frac{r}{r_0}\right)^{-\gamma}   \\
    \Sigma_{\rm p} &= \Sigma_{\rm p,0}\times\left(\frac{r}{r_0}\right)^{-\gamma}
\end{aligned}
\end{equation}
where $r$ is the distance from the star and $\Sigma_{\rm g,0}$ is the gas surface density at the reference location $r_0$. The gas disk is parameterized based on the observational constraints of MWC\,480, a typical Herbig Ae/Be star with a clear annular gap in its continuum at ${\sim}73$\,au \citep{LongEtal2018f,LiuEtal2019y}. From thermo-chemical modeling based on line emission of CO isotopes, \citet{ZhangEtal2021k} suggest a gas surface density profile of $\Sigma_{\rm g,0}=11 \rm g\,cm^{-2}$, $\gamma = 1$ at $r_0= 73$\,au. The densities of the ice-rich pebble at $r_0$ is $\Sigma_{\rm p,0}=f_{\rm m}\Sigma_{\rm g,0}$, where the mass fraction $f_{\rm m}$ depends on the volatiles species of interest in each simulation (see \Tb{abundance} for the values of each species). The stellar masses have been dynamically determined to be ${\sim}2.1\,M_\odot$ as described in \citep{TeagueEtal2021}.

We assume that the disk is isothermal in the vertical direction and employ a locally-isothermal equation of state. The ambient temperature is determined by the disk background, which is influenced by the irradiation of the host star, and the heating from the accreting planet. Specifically, the disk background temperature is described as
\begin{equation}\label{eq:Td}
    T_{\rm d} = T_0\times\left(\frac{r}{r_0}\right)^{-q}
\end{equation}
where $T_0 = 20\,K$, and $q=0.45$ according to the power-law fit of the dust temperature at the midplane \citep{SierraEtal2021}, based on the 2D (R-Z) dust density and temperature profile in the themo-chemical models in \citet{ZhangEtal2021k}. The aspect ratio of the gas disk, therefore, follows
\begin{equation}\label{eq:cs}
    h_{\rm g} \equiv \frac{H_\mathrm{g}}{r} 
    \equiv \frac{c_s}{\Omega_K r}
    = h_0\times\left(\frac{r}{r_0}\right)^{0.5-q/2}
\end{equation}
with $h_0 = 0.05$ based on the temperature profile. We note this value is inconsistent with the best-fit aspect ratio based on the SED fitting. \citet{ZhangEtal2021k} fit the vertical temperature profile at each radius directly, by matching the emission surface of CO isotope lines. Since our model focuses on the ice sublimation at the midplane, we opt for the midplane temperature in our simulations.  

As the planet is accreting, the lost gravitational energy of the accreting materials can be translated into accretion luminosity
\begin{equation}\label{eq:L_acc}
    L_{\rm acc} = \frac{GM_{\rm p}\dot{M}}{R_{\rm p}}.
\end{equation}
where $M_{\rm p}$ is the planet mass, and $R_{\rm p}$ is the radius of the planet accretion shock front, and $\dot{M}$ is the mass accretion rate of the planet. We fix $R_{\rm p} = 2R_{\rm Jupiter}$ in all of our simulations, which is the typical shock front radius of an accreting planet in the disk \citep[1.5-4\,$R_{\rm Jupiter}$, see e.g.,][and references therein]{MarleauEtal2022}. 
A detailed explanation for the prescription of planet mass $M_{\rm p}$ and accretion rate $\dot{M}$ is given in \Se{planetgrowth}. 

We utilize the accretion luminosity to prescribe the temperature surrounding the planet. 
We consider the radiation near the planet to be optically thin at infrared wavelength, which is typically the case in the planet-opened gaps at the outer regions of the protoplanetary disk. 
The temperature is thus prescribed as \citep[e.g.,][]{Rafikov2006}
\begin{equation}\label{eq:T_d}
    T = \left(\frac{L_{\rm acc}}{16\pi\sigma d^2} + T_{\rm d}^4\right)^{0.25}
\end{equation}
where $d$ is the distance to the planet.

\begin{table}[tbp]
\caption{Abundances $n({\rm X})/n({\rm H})$ for different molecules X of interest. Similar values are used in recent modeling \citep[e.g.,][]{BosmanEtal2021b,SchneiderBitsch2021a}, and consistent with cometary ices in forming protoplanetary disk midplanes \citep{DrozdovskayaEtal2016}.}
\label{tab:abundance}
\centering
\small
\begin{tabular}{l|ccc}
\hline\hline
Species & abundance & mass fraction $f_{\rm m}$ & initial phase\\
\hline
    H$_\mathrm{2}$  & 0.5                 & 0.7                 & non-condensable gas\\
    He              & 0.11                & 0.3                 & non-condensable gas\\
    CO              & $8.9\times10^{-6}$ & $1.75\times10^{-4}$ & non-condensable gas\\
\hline
    CH$_\mathrm{4}$ & $8.9\times10^{-6}$ & $10^{-4}$           & ice-rich pebbles\\
\hline
\end{tabular}
\end{table}

\subsection{Volatile sublimation}
At each grid point, volatile ice can be sublimated into vapor via a recently-developed phase change module implemented in \texttt{Athena++}  \citep{WangEtal2023}. This module solves mass transfer due to ice sublimation after each hydrodynamic step, thus enabling the self-consistent simulation of volatile transport in both ice and vapor states. The volatile follows the saturated (or equilibrium) vapor pressure, which is given by the Clausius-Clapeyron equation
\begin{equation}\label{eq:P_eq}
    P_{\rm eq} = P_{\rm eq,0}e^{-T_a/T}
\end{equation}
where $T_a$ and $P_{\rm eq,0}$ are constants depending on the molecular species. We take $T_a = 1110$\,K and $P_{\rm eq,0} = 3.67 \times 10^{9}\,\rm g\,cm^{-1}\,s^{-2}$ for methane---the major C-carrier icy volatile \citep[][where we neglect higher order correction terms]{FraySchmitt2009}. 
When the partial pressure of the volatile vapor is lower than the saturated vapor pressure, ice is assumed to be \textit{instantly} converted into vapor. In reality, the sublimation timescale of the volatile $\propto {\rm d}s/v_{\rm th}$, where ${\rm d}s$ is the depth of the volatile-rich surface layer of the pebble and $v_{\rm th}$ is the thermal velocity of vapor particles \citep[e.g.,][]{SchoonenbergOrmel2017}.
Thus the instant sublimation of ice-rich pebbles is validated for the volatiles of our interests, which are assumed covered on the surface of the pebble.
As the released vapor will mainly stay inside the gap but outside the co-orbital region, where there are few pebbles for vapor to condense on, we, for simplicity, ignore condensation in the simulations  (see  \se{limitations} for further discussion and \fg{tau_them}).

In each simulation, pebbles are assumed to be composed of refractories, which dominate by mass, and a single ice species, which can sublimate from ice to vapor phase.
Therefore, we assume that the size of pebbles is not affected by sublimation.
The abundances of all species included in our simulation are listed in \Tb{abundance}. The values are marginally the same as those taken in \citet{BosmanEtal2021b}, and consistent with cometary ices in forming protoplanetary disk midplanes \citep{DrozdovskayaEtal2016}. 
CH$_\mathrm{4}$ are ice-rich pebbles that are initially in the ice phase, in other words, the planet locates beyond their snowlines. Other materials, including CO, are assumed to be non-condensable gases throughout the simulation, which means that the planet is located inside their snowlines. For instance, as suggested in \citet{ZhangEtal2021k}, the midplane CO snowline for MWC\,480 is around 100\,au.

\begin{figure}[tb]
    \centering
    \includegraphics[width=.99\columnwidth]{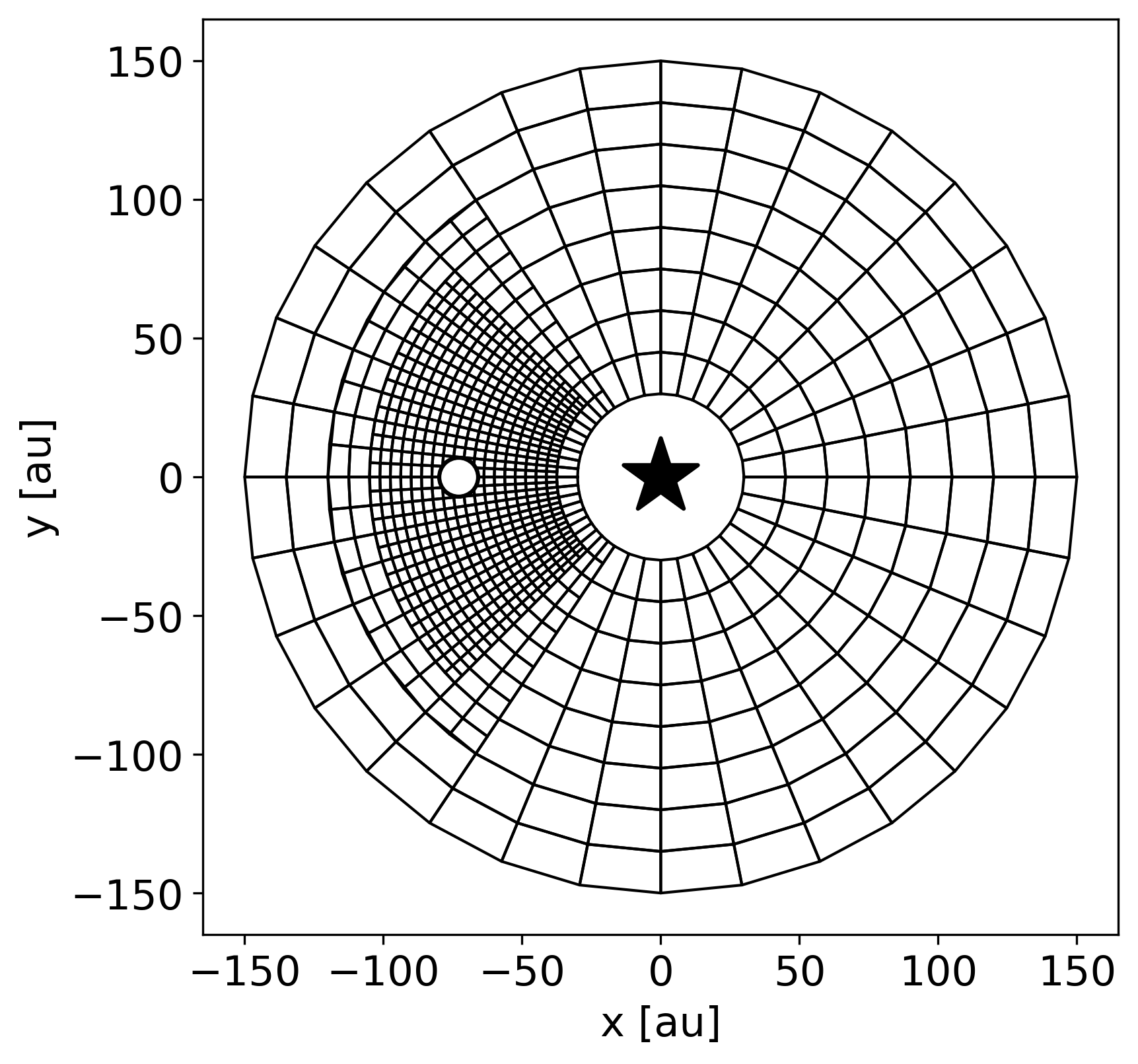}
    \caption{\label{fig:mesh_grid} Sketch of the 2D mesh grid used in our simulations. The circle represents the location of the planet (located at $r_0 = 73$\,au) and the star symbol is the central star. The radial domain ranges from 30\,au to 150\,au. The static mesh refinement level is 2, which means the smallest mesh is four times smaller than the unrefined mesh at the same location.}
\end{figure}

\begin{figure*}[tbp]
    \centering
    %\sidecaption
    \includegraphics[width=.9\textwidth]{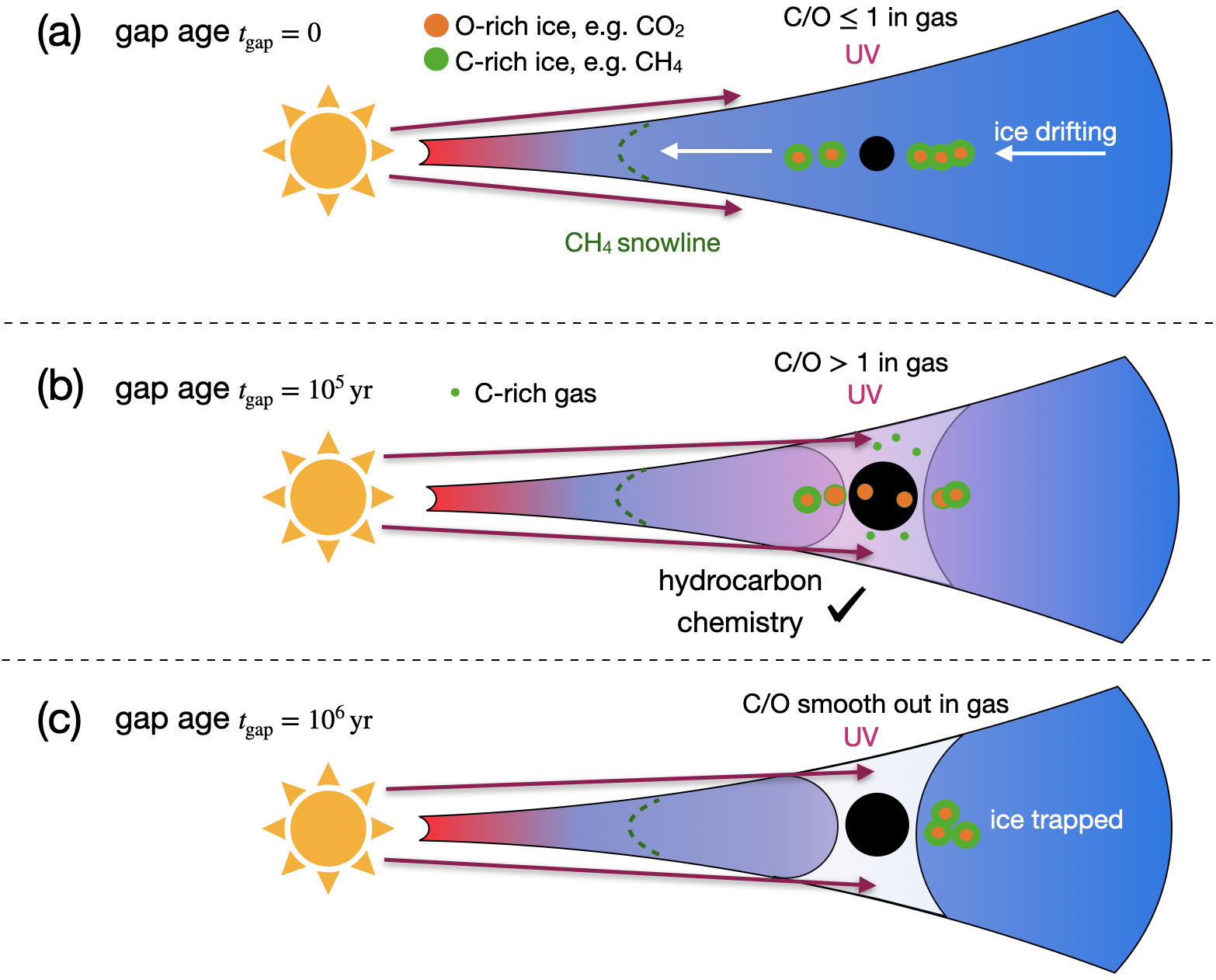}
    \caption{\label{fig:chem_sketch} Schematic of how a gap-opening planet triggers the C$_2$H ring associated with the continuum gap. {\bf (a)} Before runaway growth, a small planet is embedded inside the gaseous disk and carbon-rich ice is frozen out on pebbles. {\bf (b)} During runaway gas accretion the planet grows rapidly, consuming the pebbles in its vicinity and opening a gas gap in the disk. Since the accreting planet is hot, carbon-rich volatile ice (e.g., methane) with lower sublimation temperature is readily liberated from the ice. On the other hand, the sublimation of H$_2$O or CO$_2$ -- the major oxygen carriers -- can only be reached in the immediate vicinity of the planet (${<}R_{\rm Hill}$), by which point the material is lost to the planet. The discrepancy in sublimation temperature renders the gas C/O$>$1. In addition, stellar UV photons penetrate the gas gap. The combination of C/O$>$1 and high UV-fluxes trigger photo-chemistry reactions to form the C$_2$H ring; {\bf (c)} After the gap opens, the supply of sublimated pebbles terminates due to the gap opening. The C/O ratio is smoothed out across the disk through radial mixing by diffusion.}
\end{figure*}

\subsection{Growth of the planet}\label{sec:planetgrowth}

In our 2D simulation, we modeled the planet as a sink particle with a fixed circular orbit at a radius of $r_0$. The planet mass is initialized as $10\,M_\oplus$, a typical critical core mass for a planet starting the runaway gas accretion \citep[e.g.,][]{PerriCameron1974,Rafikov2006}, but also see recent works considering atmosphere pollution suggesting even lower values \citep{HoriIkoma2011,OrmelEtal2021}. The static mesh refinement was employed throughout the simulation to better capture the gas flow and ice sublimation in the vicinity of the planet, as depicted in \fg{mesh_grid}. The simulation domain extended radially from 30 au to 150 au, and the smallest mesh size was ${\rm d}x_{\rm min} = 0.1 R_{\rm Hill}$, where $R_{\rm Hill} \equiv (M_{\rm p}/3M_\star)^{1/3} r_0$ is the Hill radius of the planet. We used the value of $R_{\rm Hill}$ for a planet mass of $M_{\rm p} = M_0 \equiv 2.3 M_{\rm J}$, as suggested by previous studies reproducing the continuum profile of MWC\,480 \citep{LiuEtal2019y}. We tried smaller ${\rm d}x_{\rm min}$ for a convergent test, while the simulation results have no changes. A softening parameter $\epsilon = 0.1 R_{\rm Hill}$ is chosen for the gravitational potential
\begin{equation}\label{eq:grav_p}
    \Phi = - \frac{G M_{\rm p}}{\sqrt{d^2+\epsilon^2}}.
\end{equation}
The choice of $\epsilon=0.1 R_\mathrm{Hill}$ ensures that the accretion rate on the planet is high, which feature is essential to capture with our model as it heats the vicinity of the planet, allowing pebbles to sublimate. In the literature, however, a larger softening parameter is favored for 2D simulations in order to correctly capture the wave excitation and gap opening by the planet \citep[e.g.,][]{DongEtal2011,MuellerEtal2012}. Our choice for $\epsilon$ in 2D simulations could therefore expedite the gap-opening process somewhat. However, conducting a run with $\epsilon=0.6H_g$, we found little difference in the gap profile and therefore do not expect that the choice of $\epsilon$ would much affect our conclusions.

Starting with the initial planet mass of $10\,M_\oplus$, we run the simulation with the fixed-initial-mass planet for $t_\mathrm{inin} = 300$\,orbits, which is about $10^{5}$\,yr at the distance of the planet. The time is sufficiently long compared with a total disk lifetime, and enough for the disk to adapt itself to the planet embedded in the disk. We tried different $t_\mathrm{inin}$, with which our major conclusion is unaffected. After $t_\mathrm{inin}$, we assign a mass fraction of the ice-rich pebbles (\tb{abundance}), then switch on mass accretion in the simulations. We remove all pebbles entering $0.9R_{\rm Hill}$ from the simulation. The gas accretion prescription follows \citet{CridaBitsch2017} and \citet{Bergez-CasalouEtal2020}, and the accretion rate is based on the 3D isothermal shearing box simulations results from \cite{MachidaEtal2010a}, which is determined by the local gas density within $0.9R_{\rm Hill}$ of the planet
\begin{equation}\label{eq:dotM_p_1}
    \dot{M}_{\rm p} = f_{\rm A}\left<\Sigma_{\rm g}\right>_{0.9 R_{\rm Hill}} H_{\rm g}^2\Omega_K\times\min\left[0.14,\,0.83\times\left(\frac{R_{\rm Hill}}{H_{\rm g}}\right)^{\frac{9}{2}}\right].
\end{equation}
where $H_{\rm g}$ is the gas scale height. The prefactor $f_{\rm A}$ equals unity in the default setup, while we also try different values in \Se{gas_acc}. At each timestep, the accreted mass of gas is removed from the domain and added to the planet accordingly. Following \citet{Bergez-CasalouEtal2020}, the distribution of the mass removing rate $\dot{\Sigma}_{\rm g, acc}(d,t)$ reads
\begin{equation}\label{eq:dotM_p_2}
\begin{aligned}
    \dot{M}_{\rm p}
    &= \iint_{d < 0.9R_{\rm Hill}} \dot{\Sigma}_{\rm g, acc}(d,t) \,{\rm d}A   \\
    &= \iint_{d < 0.9R_{\rm Hill}} f_{\rm red}(d)\dot{\Sigma}_{\rm g}(t) \,{\rm d}A
\end{aligned}
\end{equation}
where the smooth distribution function $f_{\rm red}$ reads
\begin{equation}
    f_{\rm red} = 
    \left\{
    \begin{array}{lr}
        1 & d<0.45 R_{\rm Hill}\\
        \cos^4{\pi \left(\dfrac{d}{R_{\rm Hill}}-0.45\right)} & 0.45 R_{\rm Hill}<d<0.9 R_{\rm Hill}
    \end{array}
    \right.
\end{equation}
following \citet{RobertEtal2018}. The fraction of accreted mass increases as it gets closer to the planet, but turns flat within the innermost $0.45 R_{\rm Hill}$.  
The gas removal rate within $0.45 R_{\rm Hill}$, $\dot{\Sigma}_{\rm g}(t)$, is obtained by equating \eqs{dotM_p_1}{dotM_p_2}. In other words, the mass accretion rate of the planet is calculated through \eq{dotM_p_1}, and the accreted gas is removed from the surrounding protoplanetary disk according to \eq{dotM_p_2}.

\begin{figure*}[tbp]
    \raggedright
    \includegraphics[width=.99\textwidth]{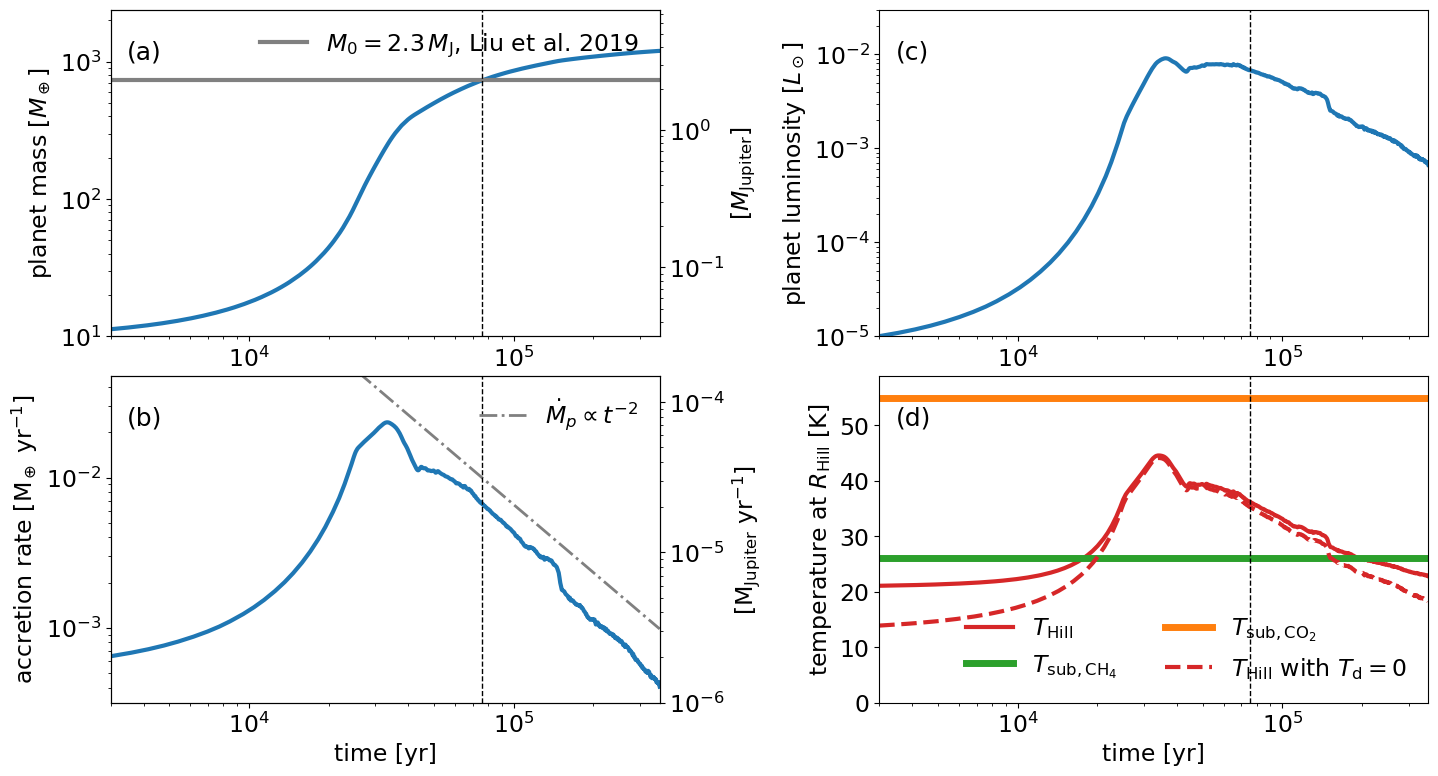}
    \caption{\label{fig:M_dotM_L_T} \textbf{(a)} Time evolution of the planet mass. The horizontal gray line indicates the planet mass $M_0 = 2.3 M_{\rm J}$ inferred by \cite{LiuEtal2019y}, which we use as a reference mass. \textbf{(b)} Time evolution of the planet's gas accretion. \textbf{(c)} Time evolution of the planet's accretion luminosity (\eq{L_acc}). \textbf{(d)} Time evolution of the disk temperature at the Hill radius. The sublimation temperature of CH$_4$ (green) and CO$_2$ (orange) are marked by horizontal lines for reference, see discussion in \se{snowline}. In each panel, the vertical dashed lines indicate the time when the mass of the planet reaches $M_0$.}
\end{figure*}

\begin{figure*}[tbp]
    \centering
    \includegraphics[width=0.88\textwidth]{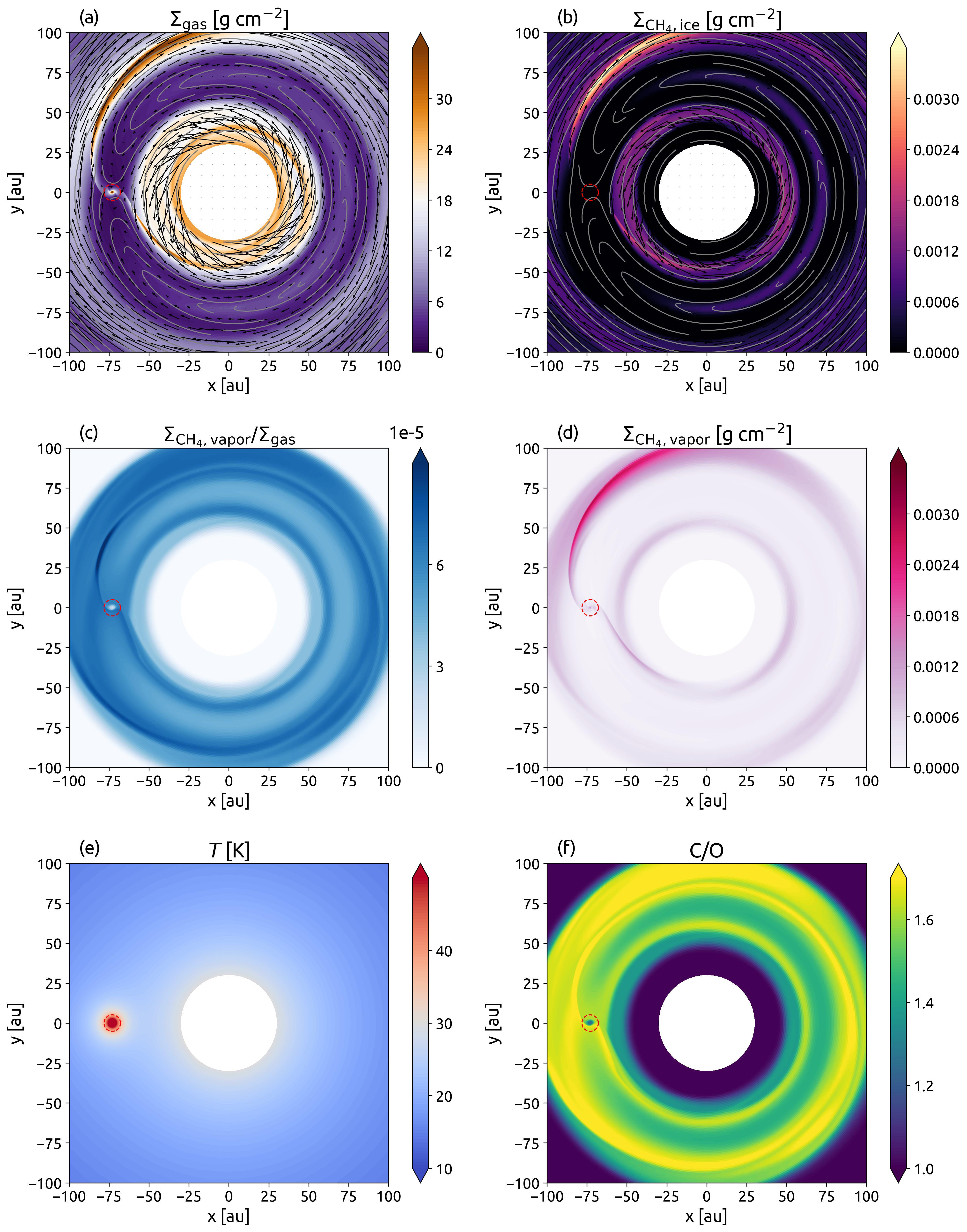}
    \caption{\label{fig:output_default_hydro} Snapshot of the fiducial model at $t=75\,751$\,yr (176 orbits), showing different components from the simulation.
    \textbf{(a)} Surface density of the gas (including vapor). Arrows indicate the velocity of the gas.
    \textbf{(b)} CH$_4$ ice surface density.
    \textbf{(c)} Surface density ratio between CH$_4$ vapor and total gas.
    \textbf{(d)} CH$_4$ sublimated vapor surface density.
    \textbf{(e)} Disk midplane temperature. 
    \textbf{(f)} C/O ratio in the gas phase. 
    In each panel, the red dashed circle indicates the Hill radius of the planet. A video showing the time evolution can be found in the online material.} 
\end{figure*}

\subsection{Sketch of the chemical evolution}\label{sec:chem_sketch}

We propose a chemical model to interpret a potential spatial correlation between the chemical ring and dust continuum profile caused by a planet exterior to the methane snowline. This model hinges on the local increase in the C/O ratio, which is attributed to localized volatile sublimation caused by the luminosity of the accreting planet. A sketch of the model can be found in \fg{chem_sketch}.

High concentrations of organics like C$_2$H and HCN indicate an active photo-chemistry, driven by high UV-fluxes and gas with a C/O ratio exceeding unity \citep{BerginEtal2016,BosmanEtal2021b}. In our model, a young giant planet is responsible for achieving these conditions. First, pebbles in the planet's orbital region are accreted by the growing protoplanet. Since the accreting planet is hot, carbon-rich volatile ices (e.g., methane) with sublimation temperatures around 25\,K are readily liberated from the ice, elevating the C/O ratio of the gas in the vicinity of the planet's orbit above unity. On the other hand, the sublimation of CO$_2$ (H$_2$O) -- the major oxygen carrier -- requires a temperature in excess of $\sim$55\,K (120\,K). Such a high temperature at 73 au can only be reached in the immediate vicinity of the planet (${<}R_{\rm Hill}$), where the water and carbon-dioxide vapor can no longer escape the planet's gravity \citep[e.g.,][]{JohansenEtal2021,BarnettCiesla2022}. In other words, the sublimation temperature difference between C-rich ices (e.g., CH$_4$, C$_2$H$_2$)and O-rich (e.g., CO$_2$, H$_2$O) renders the gas-phase C/O$>$1. Then, as the gap region becomes optically thinner due to the removal of material, stellar UV photons penetrate the gap more deeply and trigger photo-chemistry reactions. These processes may produce ring-like structures in C$_2$H \citep{BerginEtal2016}, one of the typical carbon-bearing radicals. And the more complex gas-phase formation of organic molecules compounds \citep{BastEtal2013,CalahanEtal2023}. 

\subsection{{$\rm C_2H$} chemistry}
We focus on C$_2$H rings in this work. As a radical, C$_2$H can act as a precursor for the formation of larger molecules, through photochemical reactions \cite[e.g.,][]{BastEtal2013,OebergBergin2021}. On the other hand, it is also proposed that C$_2$H can be a product of top-down photo-dissociation hydrocarbon chemistry \citep{GuzmanEtal2015}. From the observational side, \citet{BergnerEtal2019} has suggested that C$_2$H and HCN seem to share a common driver. More research is needed to fully understand the connection between C$_2$H and the formation of complex organics in these environments, where a close link between a high C/O ratio and molecular formation appears.

To estimate the column density of C$_2$H, we adopt an empirical approach instead of relying on computationally expensive full chemical network modeling methods \citep[e.g.,][]{BrudererEtal2012,DuBergin2014}. Specifically, we utilize the DALI models presented in \citet{BosmanEtal2021b}, which were calibrated to reproduce the observed C$_2$H emission levels of AS209, HD\,163296, and MWC\,480 using source-specific gas-grain thermochemical models (see their Fig. 2 and Fig. 5). 
Based on the model results of $n({\rm C_2H})$ in the three sources and the input parameters in \citet{BosmanEtal2021b}, we summarize an empirical relation (see \App{n_c2h})
\begin{equation}\label{eq:nc2h}
    \log_{10} n({\rm C_2H}) = F_C -\log_{10} n({\rm H})+2 \log_2 ({\rm C/O})
\end{equation}
where $F_C=37 \pm 1$ is a fitting constant. \Eq{nc2h} allows us to estimate the C$_2$H column density based on our simulation outputs. The dependence of $n({\rm H})$ is the result of two effects: (a) lower CO column density, therefore less self-shielding, allowing more dissociation of CO; and (b) the depletion of the small dust, which scales with the total gas density, increasing the UV penetration. The third term on the RHS of the formula directly reflects the strong dependence of C$_2$H on the carbon-to-oxygen ratio {\rm C/O}, as shown in Fig. 2 of \citet{BosmanEtal2021b}. Nevertheless, more comprehensive chemical modeling may rely on a series of additional assumptions and prescriptions, see our discussion in \Se{limitations}. 

%\hspace*{\fill} \\
Finally, we remark that although we have chosen the fiducial values of our model parameters based on the observational constraint on MWC\,480, the model we propose here is in principle applicable to any planet located exterior to its natal disk methane snowlines, see discussion in \se{snowline}. 

\section{Results}\label{sec:results}

In this section, the results of the 2D hydrodynamical simulations that implement the model introduced in \fg{chem_sketch} are presented. The fiducial model is described in \se{default}, and a parameter variation of the gas accretion rate of the planet is conducted in \se{gas_acc}.

\subsection{Fiducial model}\label{sec:default}

In our fiducial model, we include methane as the only ice-rich pebble. 
The time-dependent changes in both the mass and the accretion rate of the planet following the initiation of gas accretion are presented in panels (a) and (b) of \fg{M_dotM_L_T}.

For the first 20,000\,years (${\approx}50$\,orbits) after the onset of runaway gas accretion, the accretion rate shows an exponential increase. During this period, the gas surface density at the location of the planet does not significantly decrease. 
The reason is that the timescale for the opening of the gas gap by the planet, which is the viscous timescale over the gap width \citep{KanagawaEtal2017}
\begin{equation}
    t_{\rm gap} = 2. \times 10^3 \frac{2\pi}{\Omega_K}
    \left(\frac{M_{\rm p}/M_\star}{10^{-4}}\right)
    \left(\frac{h}{0.05}\right)^{-3.5}
    \left(\frac{\alpha}{5\times 10^{-4}}\right)^{-1.5}
\end{equation}
is longer than the orbital timescale. 
The slope of the accretion rate as a function of time slightly changes around $t = 20,000$ yr, because of the transition from Hill radius scaling towards constant prefactor 0.14 in \eq{dotM_p_1}, causing the accretion rate to flatten. Since the planet has reached a mass above $1\,M_{\rm J}$ at $t\simeq30,000$\,yr, the planet starts to sculpt the gas gap more strongly. Therefore, as the gas available around the planet decreases with the deeper and wider gap, the accretion rate of the planet goes down rapidly with a scaling of $\dot{M}_{\rm p} \propto t^{-2}$. A dot-dashed line is presented in \fg{M_dotM_L_T}(b) for reference to show the trend.

As depicted in panels (c) and (d) in \fg{M_dotM_L_T}, by using the planet's mass and gas accretion rate, we calculate the corresponding luminosity of the planet (\eq{L_acc}) and the temperature at its Hill radius
\begin{align}\label{eq:T_Hill}
    T_{\rm Hill} &= \left(\frac{L_{\rm acc}}{16\pi\sigma R_{\rm Hill}^2} + T_{\rm d}^4\right)^\frac{1}{4}
    = \left(\frac{G(3M_\star)^{2/3}M_{\rm p}^{1/3}\dot{M}_{\rm p}}{16\pi\sigma R_{\rm in} r^2} + T_{\rm d}^4\right)^\frac{1}{4} \\ \nonumber
    & = 43\,\mathrm{K} \left( \frac{M_\star}{M_\odot} \right)^\frac{1}{6}
    \left( \frac{M_{\rm p}}{M_{\rm J}} \right)^\frac{1}{12}
    \left( \frac{\dot{M}_{\rm p}}{10^{-4}\,M_{\rm J}\,\mathrm{yr}^{-1}} \right)^\frac{1}{4} \left( \frac{R_\mathrm{in}}{2\,R_{\rm J}} \right)^\frac{1}{4}
    \left( \frac{r}{73\,\mathrm{au}} \right)^{-\frac{1}{2}}
\end{align}
where $R_{\rm Hill}$ is the Hill radius of the planet. The second equation assumes that $T_d$ can be neglected. Due to the high accretion rate, the temperature inside the Hill radius is notably elevated compared to the background disk temperature, rising from $20$\,K up to approximately $45$\,K at the Hill sphere (see also \fg{output_default_hydro}(e)). Between $t=20,000$\,yr (${\approx}50$\,orbits) and $t=150,000$\,yr (${\approx}350$\,orbits), balancing the increased planet mass and the reducing amount of available gas, the accretion rate of the planet remains in the range of $10^{-5}$ to $10^{-4}\,M_{\rm J}\,\mathrm{yr}^{-1}$ during its growth, which roughly corresponds to the temperature range where $T_{\rm Hill}$ above the sublimation temperature of CH$_4$ (the horizontal green line in \fg{M_dotM_L_T}(d), see \se{snowline} and \fg{T_sub}), while is still below the sublimation temperature of CO$_2$ (the horizontal orange line in \fg{M_dotM_L_T}(d), see also \se{snowline} and \App{Orich}). In addition, we show the temperature purely contributed from the planet by taking $T_{\rm d} = 0$ in \eq{T_Hill}, which is indicated by the red line in \fg{M_dotM_L_T}(d). As the figure shows, in the region where $T>25\,$K, the temperature at Hill radius $T_{\rm Hill}$ is primarily determined by the planet's luminosity, rendering the results insensitive to the background temperature once it is below the CH$_4$ sublimation temperature. 

The simulation snapshot of the surface densities of all the components, the temperature map, and the C/O ratio map in the gas phase at $t=64,560$\,yr (176 orbits) are displayed in \fg{output_default_hydro}, along with the corresponding azimuthally averaged radial density profiles shown in \fg{output_default_rad}. At this time, the planet has reached a mass of 2.3$M_{\rm J}$, consistent with the planet mass suggested by \citet{LiuEtal2019y}, which we utilized as an educated estimate and point of reference. The Hill radius of the planet is marked by red circles in each panel. The top panel of \fg{output_default_hydro} shows the azimuthally averaged radial profile of different components. A moderate gas gap has been opened by the planet (upper left panel of \fg{output_default_hydro}). And pebbles have followed the pressure gradient, resulting in dust trapping at the outer edge of the gas gap (upper right panel of \fg{output_default_hydro}). 
The dust gap is quickly emptied, while a small fraction of pebbles remain inside the co-orbital region of the planet, where a faint arc appears. This amount of dust is also shown in the radial profile \fg{output_default_rad}. 
The opened gap allows for increased UV penetration, therefore, possibly higher temperature \citep{BroomeEtal2022}, which may help with the sublimation of the ice trapped inside the co-orbital region. Both of which is not included in our simulation, but see \Se{limitations} for more discussion.

\begin{figure}[tbp]
    \raggedright
    \includegraphics[width=.99\columnwidth]{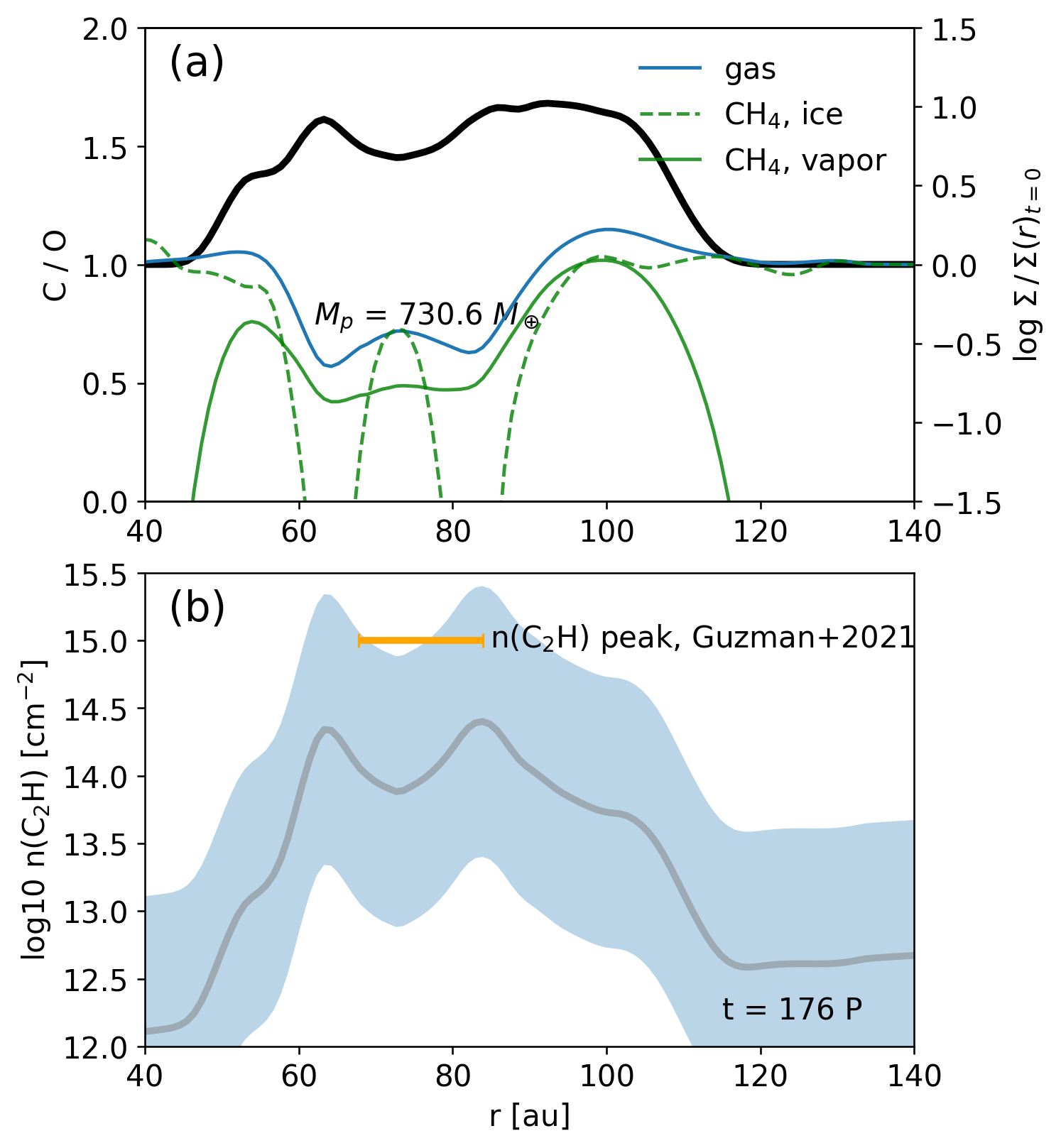}
    \caption{\label{fig:output_default_rad} Azimuthally averaged radial profiles of the fiducial model at $t=75\,751$\,yr (176 orbits).\textbf{(a)} Radial profile of total gas density (blue), CH$_4$ ice (green dashed), and CH$_4$ vapor (green solid) at 176 orbits. The gas phase C/O ratio (black) is calculated based on the abundance assumed in \tb{abundance}. \textbf{(b)} The expected C$_2$H column density based on \eq{nc2h}. The shaded region indicates the one-magnitude  uncertainty of the fitting constant $F_C$ in \eq{nc2h}. We mark the typical C$_2$H peak density (orange) from the MAPS data constraint as a reference value for illustration.}
\end{figure}

The CH$_4$ ice sublimates and transfers to vapor phase close to the planet location, where the temperature is higher compared with the disk background temperature. The majority of CH$_4$ vapor is transported away from the planet through the gas spiral arm, as shown in panels (a) and (c) of \fg{output_default_hydro}. However, a significant fraction of vapor is transported from the vicinity of the planet into the gap as well, though not exactly into the co-orbital region. When considering the CH$_4$ vapor mass fraction over the total gas density, the vapor fraction reaches a relatively high value of ${\approx}10^{-5}$ as shown in \fg{output_default_hydro}(c). 
Although the amount of CH$_4$ vapor that ends up in the planet gap region is rather low, it is nevertheless sufficient to raise the C/O ratio above unity. Using the assumed constant CO abundance of $8.9\times10^{-6}$ and the CH$_4$ vapor density, we calculate the C/O ratio in the gas phase in \fg{output_default_hydro}(f). The azimuthally averaged C/O ratio is indicated by the black solid line in the top panel of \fg{output_default_rad}. Initially, the disk is cold and there is no methane released by the planet. Thus, both carbon and oxygen in the gas phase are determined by the non-condensable CO in our model. Therefore, the gas-phase C/O ratio is unity throughout the entire simulation box at the beginning. As the CH$_4$ ---the major C-carrier volatile in our simulation--- is released from ice-rich pebbles into the gas, the C/O ratio increases accordingly.

Significant vapor release happens over a relatively short timescale at the beginning where $T_{\rm Hill}$ is above the sublimation temperature of CH$_4$, as shown in \fg{M_dotM_L_T}(d), and the gap remains shallow. In \fg{CO_t}, we show the time evolution of the radial profiles of the C/O in the gas phase. As the planet rapidly grows and gradually opens the gap, ice-rich pebbles can no longer reach the planet, and the sublimated vapor will be smeared out by radial mixing as shown in \fg{CO_t}. In reality, the CH$_4$ vapor may eventually be destroyed and converted into other species \citep[e.g.,][]{EistrupEtal2016}, or condense back to the ice. Both processes are not included in the simulations (see \se{limitations}).

\begin{figure}[tbp]
    \raggedright
    \includegraphics[width=.99\columnwidth]{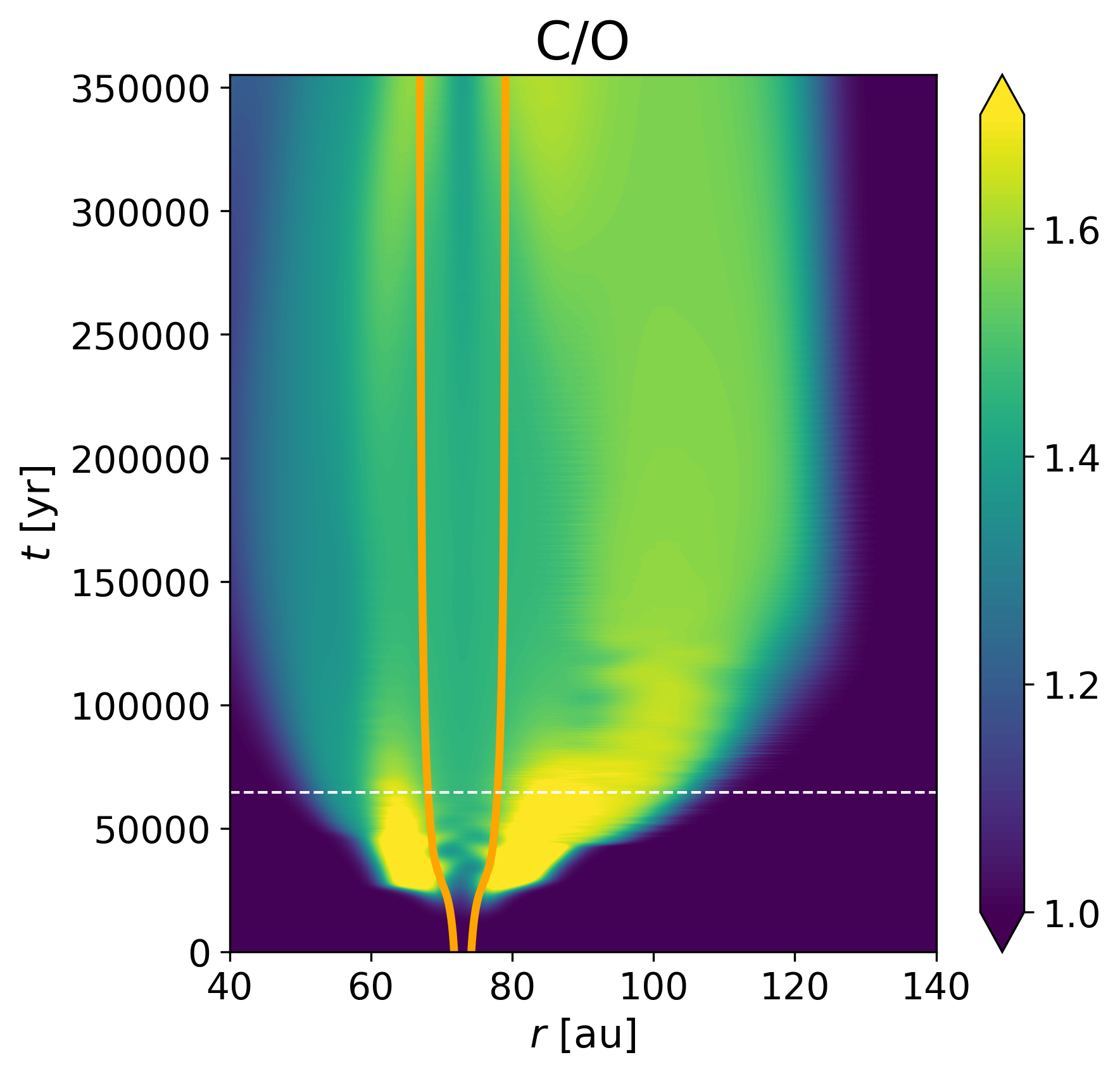}
\caption{\label{fig:CO_t} The radial profile of gas-phase C/O ratio as a function of time in the fiducial model. The horizontal white dashed line marks $t=176$\,orbits. The orange lines represent $r_0 \pm R_{\rm Hill}(t)$, where $R_{\rm Hill}(t)$ the Hill radius of the planet at each time.}
\end{figure}
Both high UV radiation and high C/O ratio are beneficial for C$_2$H formation. With the simulation outputs, we estimate C$_2$H column densities via \eq{nc2h} in the bottom panel of \fg{output_default_rad}. A clear ring profile peaks inside the dust gap. 
We mark the peak C$_2$H column densities of AS\,209, HD\,163296, and MWC\,480 estimated from fitting the C$_2$H lines hyperfine structure for reference \citep{GuzmanEtal2021}, where the observational results are from the Molecules with ALMA at Planet-forming Scales (MAPS) large program \citep{OebergEtal2021}. For all three systems, the column density of C$_2$H typically peaks ${\sim}10^{15}$\,cm$^{-2}$ at the local maximum. 
Due to the simplification of the empirical formula \eq{nc2h}, we also show the 1 magnitude uncertainty of the fitting constant $F_C$ by the gray dash region. Our model result predicts a peak C$_2$H column density value that is by-and-large consistent with the MAPS observation results. Moreover, the scenario described by the model is testable in the high spatial resolution result of ALMA, as a significant spatial correlation between the C$_2$H rings and the continuum gap is expected. We discuss this in \se{MWC480} and \se{osys}.

\begin{figure*}[tbp]
    \raggedright
    \includegraphics[width=.99\textwidth]{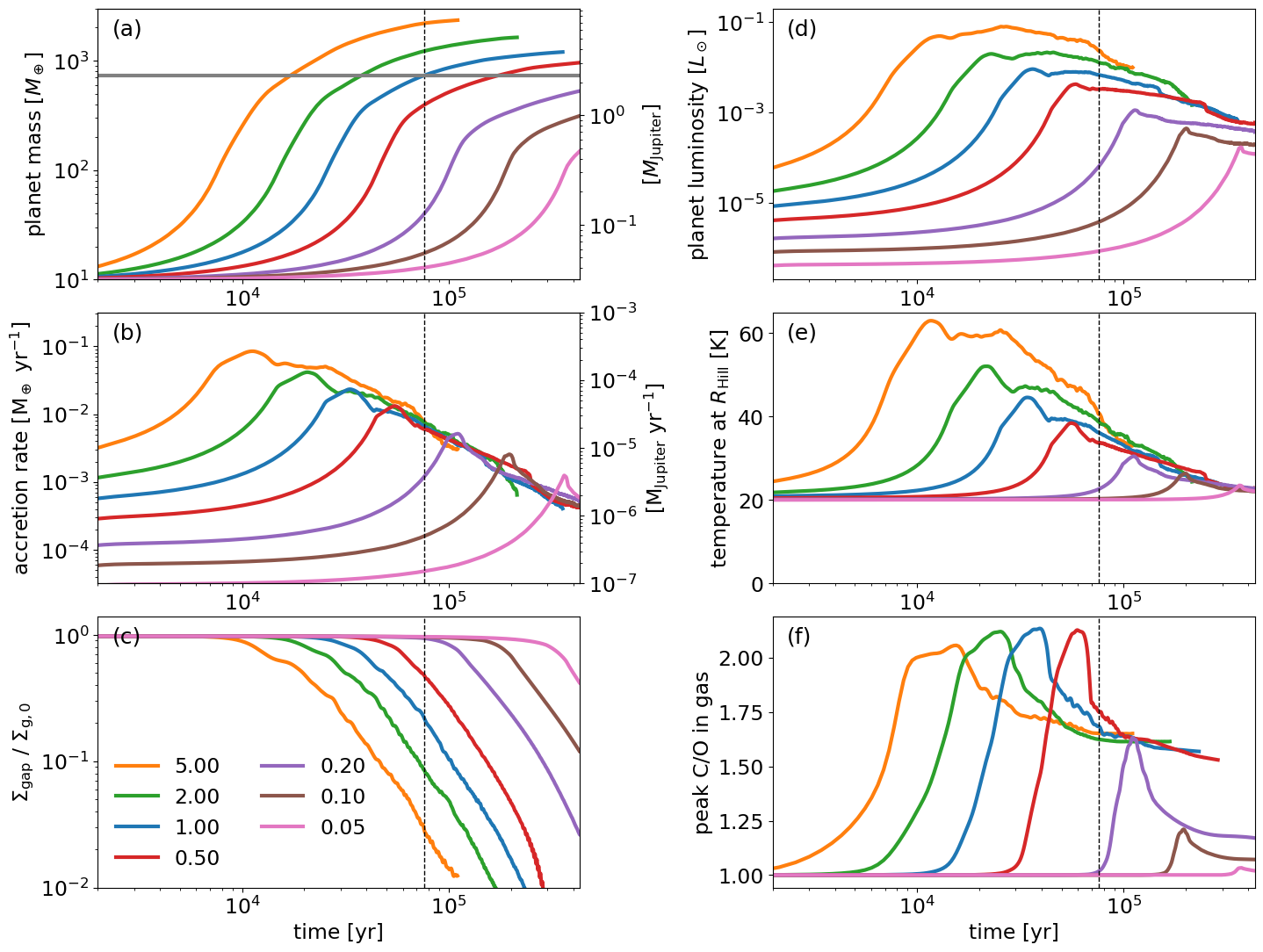}
    \caption{\label{fig:M_dotM_L_T_xA}
    \textbf{(a)} Evolution of planet mass with different gas accretion rates ($f_{\rm A}$ factor in \eq{dotM_p_1}. The horizontal gray line indicates the planet mass $M_0 = 2.3 M_{\rm J}$ inferred by \cite{LiuEtal2019y}, which we use as a reference mass. 
    \textbf{(b)} Planet's gas accretion. 
    \textbf{(c)} Gas surface density at the gap over the unperturbed surface density $\Sigma_{\rm g,0}$ as functions of time.
    \textbf{(d)} Planet luminosity (\eq{L_acc}). 
    \textbf{(e)} Disk temperature at the Hill radius. 
    \textbf{(f)} Peak gas-phase C/O ratio. 
    The prefactors $f_{\rm A}$ used in \eq{dotM_p_1} are listed in the legend.
    The vertical dashed line indicates $176$\,orbits, when the planet mass has reached $0.04, 0.05, 0.12, 1.2, 2.3, 3.8, 6.8\,M_{\rm J}$ in simulation with $f_{\rm A} = 0.05, 0.1, 0.2, 0.5, 1, 2, 5$ respectively.}
\end{figure*}

\subsection{Influence of the gas accretion rate}\label{sec:gas_acc}
The precise gas accretion rates onto planets remain an open question. Studies differ on the accretion rates attained during the runaway phase with values ranging from ${\sim}5\times$10$^{-9}$ to $\sim$3$\times10^{-4} M_{\rm J}\,\rm yr^{-1}$ \citep[][]{MachidaEtal2010a,SzulagyiEtal2014,TanigawaEtal2014,TanigawaTanaka2016,HommaEtal2020,SzulagyiEtal2022}. To investigate how the accretion rate changes our results, we vary the gas accretion formula (\eq{dotM_p_1}) by adopting different prefactor $f_{\rm A} = 0.05, 0.1, 0.2, 0.5, 1, 2, 5$ in the equation.

The planet masses and accretion rates are shown in \fg{M_dotM_L_T_xA}(a) and (b) separately, with different colors indicating different prefactors $f_{\rm A}$. We also show the gap depth (the gas surface density at the planet-opened gap $\Sigma_{\rm gap}$ over the unperturbed surface density $\Sigma_{\rm g,0}$) as functions of time in \fg{M_dotM_L_T_xA}(c). At early times, when the gap remains shallow ($\Sigma_{\rm gap}/\Sigma_{\rm g,0} \sim 1$), all simulations show an exponentially growing accretion rate. The accretion rate increases with larger $f_{\rm A}$, resulting in a more massive planet within the same period after the onset of gas accretion. At the same time, the simulations with higher accretion rates open the gas gap faster, through a combination of increased accretion and stronger torques from the more massive planet pushing the gas away. As a result, it reaches the maximum accretion rate at an earlier turning point. In \fg{M_dotM_L_T_xA}(d) and (e), the corresponding planet luminosity and temperature at the Hill radius $T_{\rm Hill}$ at each run are calculated. 
The increased mass and accretion rate leads to a higher temperature around the planet, which reaches $50$\,K in simulation with $f_{\rm A} = 2$ and $60$\,K in simulation with $f_{\rm A} = 5$, causing more sublimation of ice-rich pebbles in the planet's vicinity within the same time. Since the accretion rates decay in a similar fashion, $\dot{M}_{\rm p} \propto t^{-2}$, after passing the peak, a higher accretion rate peak means that the temperature around the Hill radius of the planet remains higher than the CH$_4$ sublimation temperature for a longer time.

We present the maximum gas-phase C/O ratio versus time in \fg{M_dotM_L_T_xA}(f). Initially, when the planet is accreting slowly, carbon in the gas phase is dominated by CO, the C/O ratio is unity, and UV photons cannot penetrate the shallow gap. These conditions are unfavorable for the formation of C$_2$H rings. In simulations with faster accretion rates (more massive planets), the C/O in the gas phase rises faster and earlier. Yet, the maximum C/O ratio in the gas phase is limited by the (fixed) amount of carbon ice in the pebble reservoir. 
By virtue of the assumption that the total (gas+dust) CO and CH$_4$ abundances are comparable, the maximum C/O ratio becomes ${\sim}2$ when methane ice fully sublimates into vapor. 
Additionally, a more massive planet opens a deeper gap, allowing more UV radiation penetration and promoting chemistry, which both facilitate C$_2$H formation. Hence, a higher gas accretion rate is preferred to generate C$_2$H rings.

Conversely, in simulations with $f_{\rm A} \leq 0.2$, the required growth timescale of the planet is much longer. The planet's mass grows slowly, resulting in a relatively low accreting luminosity, even at the accretion rate peak (\fg{M_dotM_L_T_xA}(e)). Consequently, only a fraction of methane ice sublimates to vapor, leading to a peak C/O ratio of only 1.6 in the run with $f_{\rm A} = 0.2$, and even lower, barely exceeding unity, in the runs with $f_{\rm A} = 0.1$ and $f_{\rm A} = 0.05$.
As the gap becomes deeper, pebbles are pushed away from the planet and trapped at the outer edge of the gap. In all simulations, the gas-phase C/O ratio around the continuum gap is then gradually smoothed out due to the viscous evolution of the gas, as shown in \fg{CO_t}, and the peak value decreases. 

\begin{figure}[tbp]
    \centering
    \includegraphics[width=.99\columnwidth]{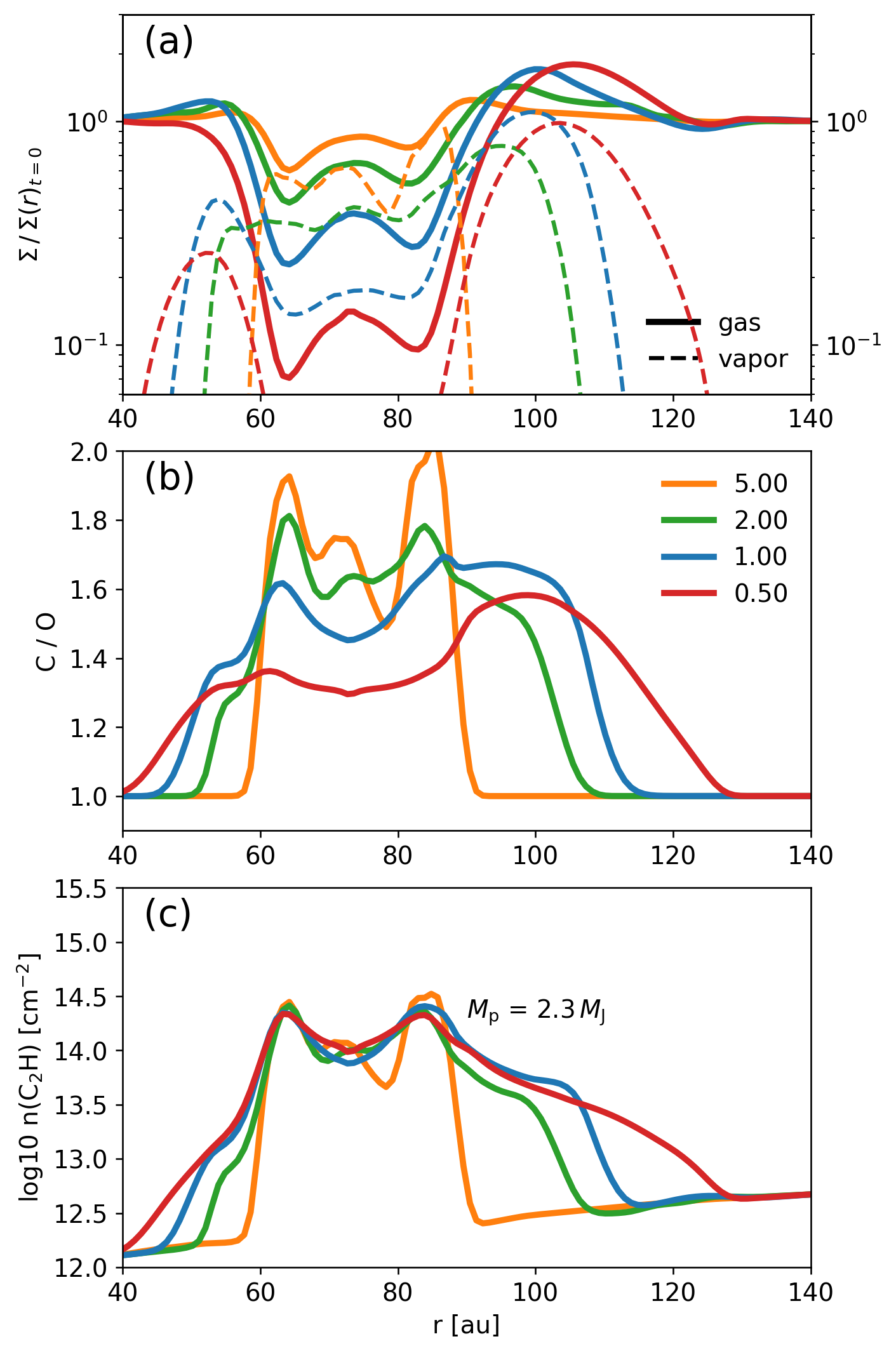}
    \caption{\label{fig:output_m23_xA} Azimuthally averaged radial profiles of simulations with different accretion rates $f_{\rm A}\geq 0.5$ at $M_{\rm p}=2.3$\,$M_{\rm J}$. \textbf{(a)} Total gas density (solid), CH$_4$ vapor density (dashed) at the indicated planet mass over the unperturbed density gas and pebble profile \eq{Sig}. \textbf{(b)} C/O ratio in the gas (midplane) in our simulations based on the abundance assumed in \tb{abundance}. \textbf{(c)} Expected C$_2$H column density.}
\end{figure}

\begin{figure}[tbp]
    \centering
    \includegraphics[width=.99\columnwidth]{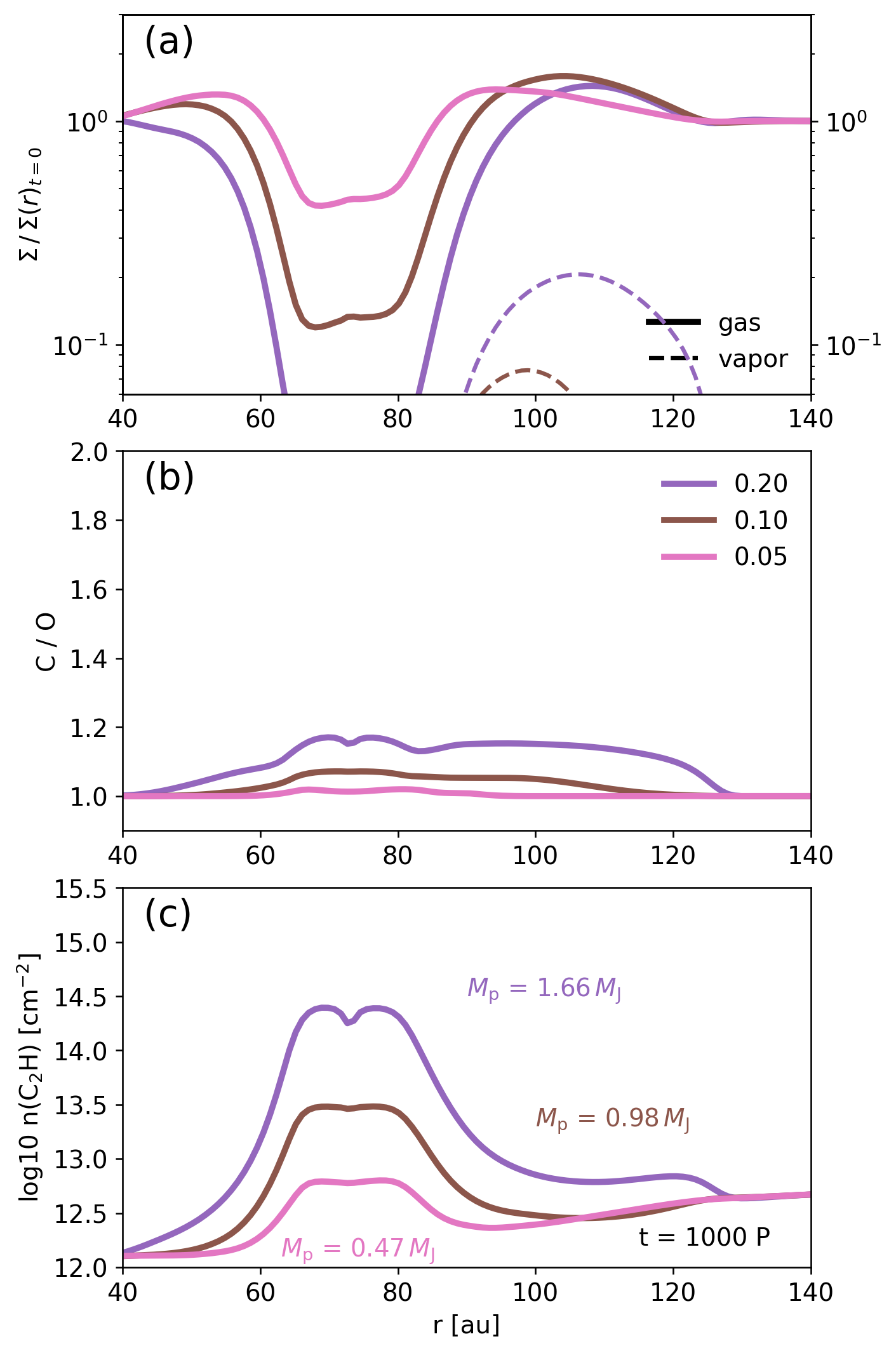}
    \caption{\label{fig:output_p1000_xA} Azimuthally averaged radial profiles of simulations with different accretion rates $f_{\rm A}\leq 0.2$ at $t=1000$\,orbits. Panels are the same as \fg{output_m23_xA}. The amount of the released carbon is relatively low compared with the high accretion rate runs. High C$_2$H column density might still be reached when the gap is very deep.}
\end{figure}

We then compare simulations with different prefactors $f_{\rm A}\geq 0.5$ at various time steps: $t=40$, 88, 176, and 402. These timesteps correspond to the point when the mass of the planet has just reached $2.3\,M_{\rm J}$ in each run. \Fg{output_m23_xA} displays the radial profiles of gas, ice, and vapor separately, as well as the correspondingly estimated C$_2$H column density profiles. 
The gas profile and C/O ratio exhibit significant differences among the simulations, but no apparent difference in C$_2$H production is observed in simulations with $f_{\rm A} > 0.5$. Planets with higher accretion rates reach $M_0$ more quickly and therefore have shallower gaps. Shallower gas gaps mean less UV penetration, which disfavors C$_2$H formation. However, the higher luminosity and shallower gap allow for more methane to be released from ice-rich pebbles via sublimation, resulting in a higher C/O ratio. These two effects balance each other, leading to similar C$_2$H column densities across different accretion rate setups. As a result, inferring a planet's mass or luminosity from the radial profile of C$_2$H can be challenging.

For the other low accretion rate runs with $f_{\rm A} \leq 0.2$, the planet mass has not reached $2.3\,M_{\rm J}$ at the end of the simulation time of 1000 orbits, the result at the end of the simulation are shown in \fg{output_p1000_xA}, where the planet masses are $1.66\,M_{\rm J}$, $0.98\,M_{\rm J}$ and $0.47\,M_{\rm J}$ in the simulation with $f_{\rm A} = 0.2$, 0.1, 0.05, respectively. 
Once the planet opens a sufficient gap, the accretion rate passes its peak turning point, as shown in \fg{M_dotM_L_T_xA}(b), and the temperature at the Hill radius follows suit, as seen in \fg{M_dotM_L_T_xA}(e). As the temperature around the planet cools down and the gas gap prevents further pebble supply, running the simulation further does not aid in methane release after the accretion rate peak is passed, as shown in \fg{M_dotM_L_T_xA}(f). All of our simulations have passed the peak accretion rate/temperature point.

As discussed above and shown in \fg{output_p1000_xA}, if the accretion rate drops below $f_{\rm A} \leq 0.2$, the results will be qualitatively different. Specifically, if the peak luminosity of the planet is too small, the disk will remain too cold to release any methane throughout the gas accretion period (\fg{M_dotM_L_T_xA}(f)).
Although the increased UV penetration might be helpful with increasing the C$_2$H density inside the gap, the absolute column density of C$_2$H would still be insufficient to account for the observed values in ALMA observations, unless a very deep gap is opened (gap depth $\Sigma_{\rm gap}/ \Sigma_{\rm g,0} \ll 0.01$), as seen in \fg{output_p1000_xA} for the simulation with $f_{\rm A} = 0.2$. We acknowledge that this result is sensitive to the empirical formula used in \eq{nc2h}, and high C/O ratios are essential for reproducing the observed value in the literature with more detailed chemical networks \citep{BosmanEtal2021a,BosmanEtal2021b}. The possibility that C$_2$H formation is mainly due to high UV penetration is proposed by recent work \citep[e.g.,][]{CalahanEtal2023}, which should be particularly interesting in transitional disks and some deep gas gaps. Since the majority of gas gaps found in, e.g., MAPS samples are relatively shallow \citep[gaps depth $\Sigma_{\rm gap}/ \Sigma_{\rm g,0} \gtrsim 0.1$, Table 4 in ][]{ZhangEtal2021k}, this routine is not the primary focus of this work.

Thus, a dichotomy exists between "hot" and "cold" planets, determined by the maximum accretion rate the planet can reach. For the disk setup used in this study, a planet accretion rate above ${\sim}10^{-5} M_{\rm J}\,\rm yr^{-1}$ is necessary to create a prominent C$_2$H ring whose peak intensity is comparable to those detected in MAPS \citep{GuzmanEtal2021}. This threshold is only weakly dependent on the planet mass $M_{\rm p}$, as \eq{T_Hill} shows that $T_{\rm Hill} \propto M_{\rm p}^{1/12}\dot{M}_{\rm p}^{1/4}$. Besides, planet mass is always around $M_{\rm J}$ at the accretion rate peak. Therefore, the maximum accretion rate that a planet can achieve will primarily determine the evolution of volatiles in its vicinity. Conversely, the observed spatial correlation between a C$_2$H ring and a continuum ring can, in principle, constrain the accretion history of the putative planet.

\section{Discussion}\label{sec:discussion}
Our findings suggest that the formation of a giant planet could have a substantial impact on the chemistry of protoplanetary disks by locally altering the elemental abundance in the gas and ice phases. 
In the following, we list the main caveats of this work and we broadly discuss some applications and implications of our model.

\subsection{Limitations}\label{sec:limitations}
One major limitation of our study is that the hydrodynamical simulation is constrained within a 2D framework. To fully capture the complexity of planet accretion, it necessitates the use of high-resolution, 3D hydrodynamical simulations. Previous studies have demonstrated that the accretion rate and flow pattern may vary significantly in different model setups for both gas \citep[e.g.,][]{MachidaEtal2010a,TanigawaEtal2012,FungEtal2019} and solid materials \citep[e.g.,][]{TanigawaEtal2014,HommaEtal2020,SzulagyiEtal2022}. In particular, some major differences can be caused by the lack of meridional flow, which can only be probed in 3D simulation \citep[e.g.,][]{SzulagyiEtal2014,FungChiang2016,TeagueEtal2019}. The meridional flow would trigger strong vertical and radial mixing, further complicating the elemental mixing and chemistry. Nevertheless, the computational cost of 3D high-resolution hydrodynamical simulations of gas accretion is substantial, hindering our ability to study the evolution of these systems using current computing resources.

In addition, the lack of a vertical dimension in the simulations limits our ability to account for the stellar UV radiation, and renders the related disk irradiation temperature and UV photo-chemistry uncertain. As recent works have shown, the planet-opened gap may also become warmer due to higher irradiation from the star \citep{BoothEtal2021a,PirovanoEtal2022,BroomeEtal2022}. The parameterized temperature used in our simulation might only be a reference lower limit.
Moreover, the growth and settling of pebbles can affect the vertical transport of ice reservoirs and the depletion of small dust in the upper layer of the disk, which can also potentially elevate the C/O ratio of the gas and increase the UV penetration, as demonstrated by \cite{VanClepperEtal2022}.
Due to the above reasons, we opted for an empirical C$_2$H column density estimation in this work (see \Se{chem_sketch} and \App{n_c2h}). The absence of chemical reactions does not impact our conclusion regarding C/O elevation as long as the reaction product remains in the gas phase.
To better reproduce the ALMA observational results and understand the astrochemical evolution of protoplanetary disks, future follow-up tests of our model featuring simultaneous hydrodynamic simulation and chemical network are worthwhile \citep[e.g., see some pioneer efforts, ][]{GongEtal2017,GongEtal2023,IleeEtal2017,BoothIlee2019,KrijtEtal2020,BergnerCiesla2021,HuEtal2023}.

Another caveat is that we do not include condensation of vapor in our simulation. 
Based on the simulation output in our fiducial model, we calculate the condensation timescales for methane \hjch{, where we assume spherical pebbles and one typical particle size at each grid point. The particles size
\begin{equation}
    s_p = \frac{2}{\pi} \frac{{\rm St}\Sigma_{\rm g}}{\rho_\bullet}
\end{equation}
is calculate by assuming a value for the stokes number $\rm St = 0.01$ and internal density $\rho_\bullet = 1.67 \rm g\,cm^{-3}$ \citep{BirnstielEtal2018}. Therefore, the typical particle mass follows $m_p = (4/3)\pi s_p^3 \rho_\bullet$. The}{vapor back to} particle\hjadd{s with} size \hjch{is fixed at}{of} ${\sim}$0.5\,mm\hjch{at the planet orbit }{, the pebble size which is used in the hydrodynamical simulations (see \Se{model}).} The condensation timescale reads \citep[e.g.,][]{SchoonenbergOrmel2017}
\begin{equation}\label{eq:t_con}
    \tau_{\rm con} 
    \equiv \frac{\Sigma_{\rm vapor}}{\dot{\Sigma}_{\rm vapor\rightarrow ice}}
    =\frac{1}{8} \sqrt{\frac{\mu_Z}{k_BT}} \frac{m_p}{s_p^2} \frac{H_{\rm g}}{\Sigma_{\rm pebble}}
\end{equation}
where $\mu_Z = 16 m_{\rm H}$ is the mean molecular weight of methane vapor, and $m_{\rm H}$ is the proton mass, $\Sigma_{\rm pebble}$ is the surface density of the pebbles and $H_{\rm g}$ is the gas scale height\hjadd{, and the typical particle mass follows $m_p = (4/3)\pi s_p^3 \rho_\bullet$}. A map of the condensation timescale based on our simulation outputs is shown in \fg{tau_them}\,(a). \hjadd{In regions of high pebble density, outside the gap region, the condensation timescale is shorter than the orbital timescale. In addition, the co-orbital region of the planet, where a moderate amount of pebbles linger, is characterized by short condensation timescales. Overall, however, d}ue to the depletion of solids, the condensation timescale in the gap region is much longer than the orbital period, with maximum values of ${\sim}10^5$yr. 
\hjadd{In particular, the C/O ratio is higher in the region with long condensation timescales, as illustrated by the green contour in \fg{tau_them}, which indicates C/O = 1.5. Regions with condensation timescales exceeding 250 orbits (white) are associated with C/O>1.5. Therefore, we conclude that accounting for the condensation of methane in the simulations will not prevent high C/O ratios in the gap, and therefore our major conclusions.}

Similarly, the sublimation timescale reads (see a similar formula for the desorption timescale in \citet{PisoEtal2015b})
\begin{equation}\label{eq:t_sub}
    \tau_{\rm sub} 
    \equiv \frac{\Sigma_{\rm ice}}{\dot{\Sigma}_{\rm ice\rightarrow vapor}}
    =  \frac{1}{8\sqrt{2\pi}} \sqrt{\frac{k_BT}{\mu_Z}} \frac{m_p}{s_p^2} \frac{1}{P_{\rm eq}}
\end{equation}
which strongly depends on the equilibrium pressure $P_{\rm eq}$ (\eq{P_eq}). In \fg{tau_them}\,(b), the sublimation timescale displays a sharp transition around the planet Hill radius, within which the sublimation timescale becomes ${\ll}1$\,orbit, much shorter than the local spreading time, validating our assumption of "instantaneous" sublimation. The quick sublimation around the planet and the long condensation timescale in the gap indicate that the methane vapor will first spread through the co-orbital region (which happens on a synodical timescale of several orbital periods) before the vapor has a chance to condense back the "cold" pebble at the midplane.

In addition, the reduced dust optical depth in the gap enables radiation to penetrate deeper into the disk, more effectively warming the gap \citep[e.g., ][]{CalahanEtal2021,BroomeEtal2022}. Finally, through strong vertical transport such as meridional flows) \citep{TanigawaEtal2012,MorbidelliEtal2014}, the released vapor may reach the upper layer of the disk, where temperatures are known to be higher \citep[e.g.,][]{D'AlessioEtal1998}. Therefore, the recondensation of CH$_4$ vapor onto settled pebbles is likely to be ineffective in the gap region. It is worth noting that our results are not \hjch{affected by}{sensitive to} vapor condensation onto grains of size $\ll$ mm. \hjch{Even though the}{Assuming} condensation timescale \hjch{could be}{are} short \hjch{, those small grains are well}{and grains perfectly} coupled with the gas\footnote{Dust coagulation may be ineffective inside the gap due to the low density. However, small particles can still be aerodynamically large, preventing vertical diffusion and allowing for sweep-up by pebbles. 
A detailed model of the balance between vertical transport and the thermal process is beyond the scope of this work \citep[but see][for example]{VanClepperEtal2022}.}
\hjrem{Therefore, any} the ice re-condensed on the surface of these small grains would \hjch{reach the upper layers of the disk, where it is}{nevertheless be} exposed to sufficiently high doses of UV photons \hjadd{at the surface of the disk} \hjch{and higher temperatures to release}{, releasing the} carbon \hjadd{back} into the gas. 

\begin{figure*}[tbp]
\centering  
%\subfiguretopcaptrue
\mbox{}\hfill
\subfigure
{\label{fig:t_con}\includegraphics[width=0.39\textwidth]{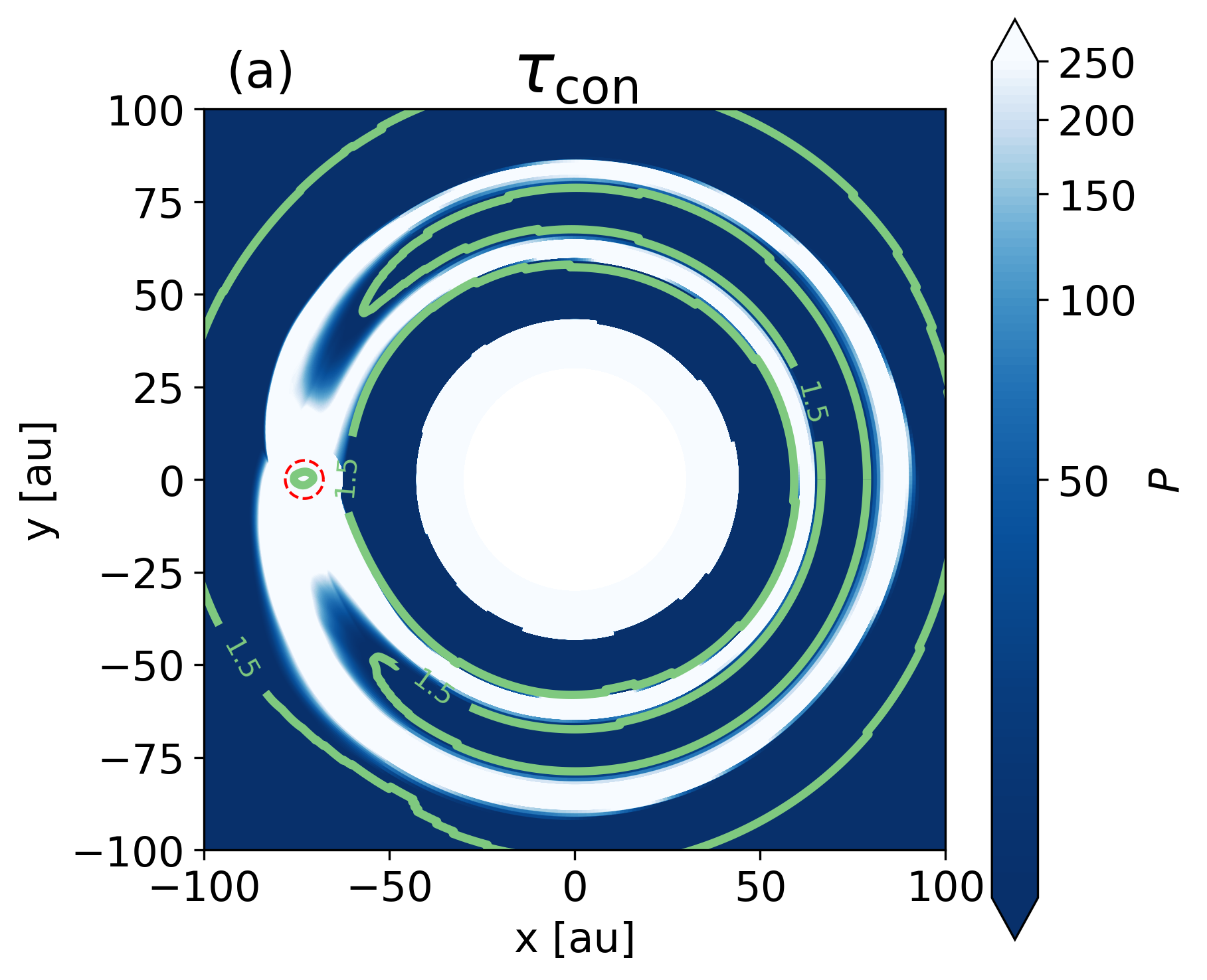}}\hfill
\subfigure
{\label{fig:t_sub}\includegraphics[width=0.39\textwidth]{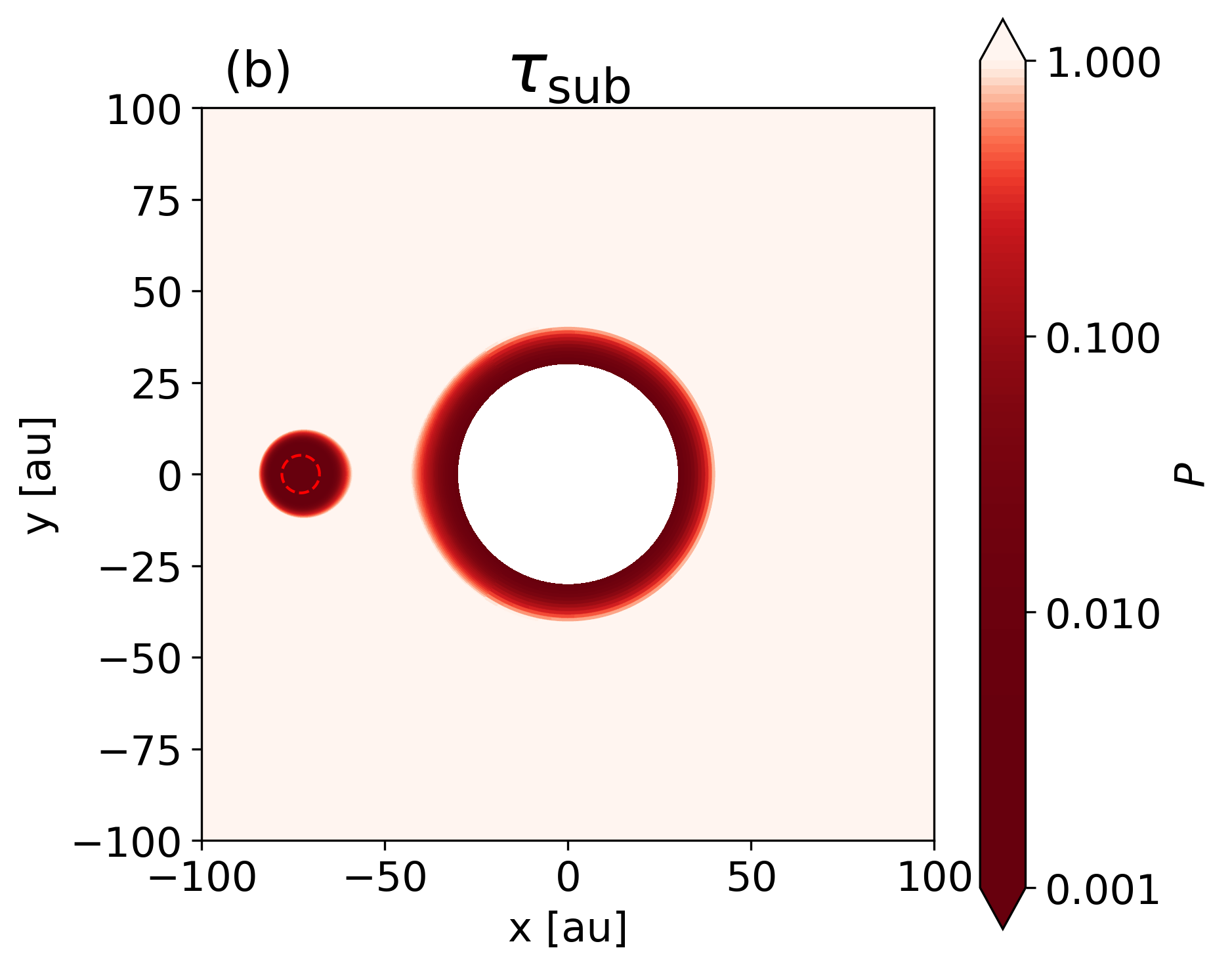}}\hfill
\hfill
%\subfiguretopcapfalse
\vspace{0. cm}
\caption{Maps of \textbf{(a)} condensation timescale and \textbf{(b)} sublimation timescale of methane ice in the default simulation at $t=75751$\,yr (176 orbits), expressed in units of the orbital period. With the exception of the co-orbital region, the condensation timescale of methane in the gap region is much longer than the synodical timescale ${\sim}10$ orbits. This allows the gas to spread azimuthally. Since moderate amounts of pebbles may remain inside the co-orbital region, the condensation timescale could still be low there. Yet, the C/O ratio is higher in the region with long condensation timescales as shown in \fg{output_default_hydro}~(f), see also the green contour which indicates the C/O = 1.5 boundary. Regions with condensation timescales exceeding 250 orbits (white) are associated with C/O>1.5. The red dashed circles indicate the Hill radius of the planet.}
\label{fig:tau_them}
\end{figure*}

On the other hand, condensation is bound to happen in the colder regions exterior to the continuum gap, where densities are high and temperatures low. As we do not include condensation \hjadd{processes} in our simulation, the methane vapor \hjch{that has been transported towards}{seen beyond} 100 au and the corresponding plateau in the C/O ratio \hjrem{profile near 100\,au} are artificial (see, e.g., \fg{output_default_rad}).
Hence, the peak in CH$_4$ vapor at 100\, au seen in \fg{output_default_rad}a and the corresponding shoulder seen in \fg{output_default_rad}(b) will, in reality, be absent. We have verified that the conclusions of this study, in particular, the peak C$_2$H are unaffected by its presence, however.

Finally, the model neglects any potential migration of the planet \citep[see reviews of ][]{KleyNelson2012,PaardekooperEtal2022}. Given the short chemical \citep{BosmanEtal2021a} and sublimation (desorption) timescale \citep{PisoEtal2015b} of $10^{4}$\,yr in the disk region of our interests, as well the short gas accretion windows shown in our simulation, our results should be unaffected by the planet migration happening over a timescale longer. Once the planet opens the gas gap, the migration speed of the planet is slower \citep[e.g.,][]{KanagawaEtal2018}, which further validates our assumption of a slow migration.

\begin{figure}[tbp]
    \centering
    \includegraphics[width=.99\columnwidth]{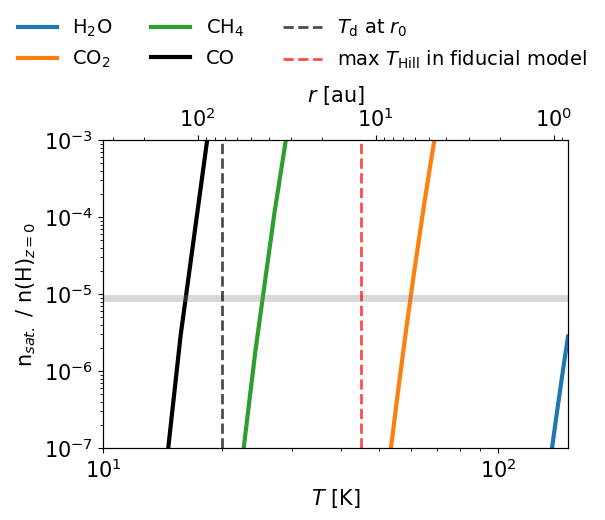}
    \caption{Location of snowlines at midplane for molecules of interest in this paper. Based on the midplane gas density at each radius, equating the partial pressure with \eq{P_eq}, one can derive the saturation abundance of molecules (solid lines). The CO and CH$_4$ abundance used in this work ($8.9\times 10^{-4}$, gray horizontal line) is given as a reference.}
    \label{fig:T_sub}
\end{figure}

\begin{figure*}[tbp]
    \centering
    \includegraphics[width=.9\textwidth]{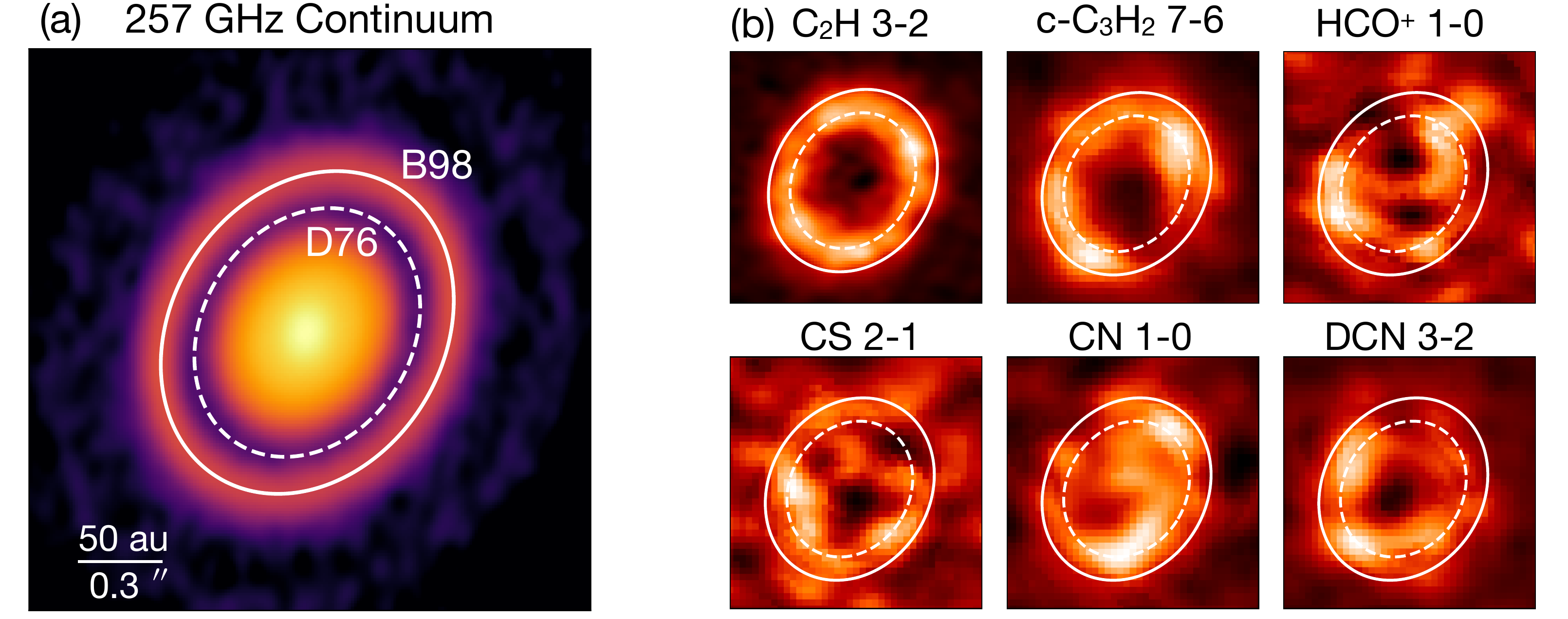}
    \caption{\label{fig:MWC480_obs} ALMA Observations of the MWC\,480 disk. {\bf (a):} 257 GHz mm dust continuum emission \citep{SierraEtal2021}. {\bf (b):} Line emission moment 0 maps \citep{LawEtal2021a} of six molecular transitions (selected for illustrative clarity; in total over 10 lines feature similar ring-like structures). Names of lines are labeled above each panel. The dashed ellipse marks the D76 continuum gap, and the solid ellipse marks the B98 continuum ring on the outside in all panels.}
\end{figure*}

\begin{figure*}[tbp]
    \centering
    \sidecaption
    \includegraphics[width=.69\textwidth]{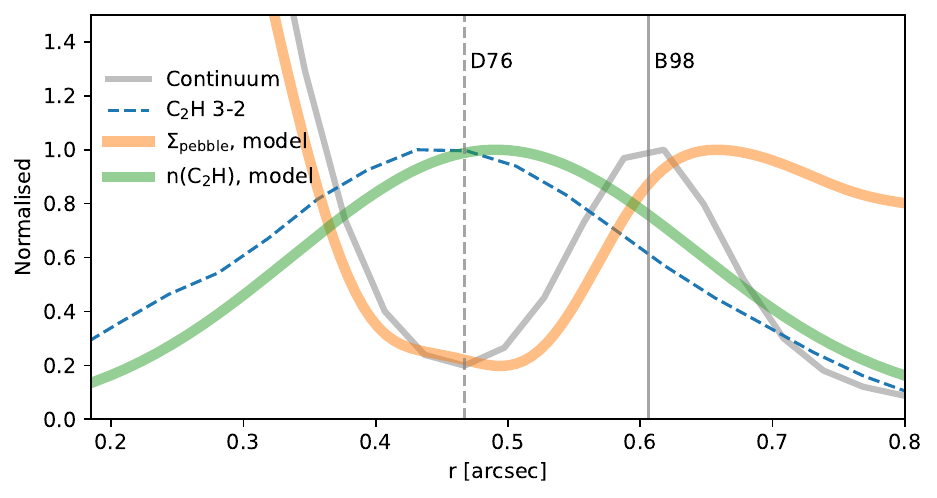}
    \caption{\label{fig:MWC480_rad} Azimuthally-averaged continuum emission profile (gray, normalized to unity at B98) and intensity profiles of C$_2$H 3-2 (blue, normalized to unity at the peak). The normalized pebble surface densities and C$_2$H column densities profile at $t = 176$\,orbits in the default run are shown in orange and green for comparison}, which are convolved by 0.1\,arcsec and 0.15\,arcsec beams used in MAPS observation separately. The dashed and solid vertical lines coincide with the D76 gap and the B98 ring in the continuum, respectively.\vspace{3mm}
    \vspace{-3mm}
\end{figure*}

\subsection{Location of the planet relative to different snowlines}\label{sec:snowline}
In our scenario, the ability for an accreting planet to elevate the C/O ratio rest on two conditions: (1) the planet should be located outside the disk CH$_4$ snowline; (2) the planet's accretion luminosity should be such to sublimate the CH$_4$, but not O-carriers like CO$_2$ or H$_2$O. To understand how general these conditions are, we plot in \fg{T_sub} the saturation profiles of said molecules for the disk model adopted in this work. In \fg{T_sub} the solid curves show the gas-phase abundance ($n_i/n({\rm H})$) one obtains by assuming complete saturation, i.e., when the partial pressure equals the equilibrium pressure of the vapor, $P_Z = P_\mathrm{eq}$, where the equilibrium pressure for different species are taken from \citet{{FraySchmitt2009}} as described in \App{thermo}. In other words, the curves give the maximum vapor abundance for each species at a given radius or corresponding temperature. It shows that at the initial background temperature of 20\,K (73\,au), all CO is in the gas phase, whereas CO$_2$ sublimation remains insignificant even at the point where the highest $T_{\rm Hill}$ is reached in the fiducial model.

In particular, a local C/O$>$1 gas enhancement will always appear whenever the planet is located beyond the methane snowline  (${\approx}40$\,au). 
As was shown in \eq{T_Hill} and \fg{M_dotM_L_T}(d), the peak temperature is determined by the accretion luminosity of the planet, independent of the background disk temperature. Furthermore, The presence of the CO snowline does not influence the outcome of a C/O$>$1 gas.

On the other hand, accreting planets located between snowlines of species with higher sublimation temperatures may leave a different chemical footprint. For instance, an accreting planet located between the methane and carbon dioxide snowlines may, in turn, lower the C/O ratio in its vicinity by unleashing more oxygen than carbon through the local sublimation of carbon dioxide
The basic tenet of this paper -- an accreting planet can locally alter the gas chemistry by selective sublimation of ices  --  applies generally.

\subsection{Exoplanet atmosphere and disk chemistry link}
The relationship between the astrochemistry in protoplanetary disks and planetary atmospheres was first highlighted by \citet{OebergEtal2011}. The authors propose that examining the C/O ratio in planetary atmospheres may allow us to trace the location where the majority of the planetary gas accretion happened, as the C/O of both gas and solid phases varies as ladders with increasing radius in the protoplanetary disk. This picture, in which giant planets inherited their atmosphere directly from the gaseous protoplanetary disk, has been actively developed in the past decade from both theoretical and observational aspects \citep[e.g., ][]{EistrupEtal2018,OebergWordsworth2019,BoothIlee2019,CridlandEtal2019,CridlandEtal2020b,GRAVITYCollaborationEtal2020,Molli`ereEtal2022}. 

Conversely, our results show that giant planet formation may have significant local impacts on the astrochemistry of the disks. The gap opening and accretion of the planet can locally alter the chemical composition of the disks. Specifically, with a planet located outside the methane snowline, the water and carbon dioxide ice component of the accreted pebbles end up in the atmosphere of the planet, which would therefore become O-rich. In contrast, CH$_4$ sublimates from pebbles to vapor, which renders the gap region C-rich. 
Our results indicate that giant planets actively shape their environment not only by opening gaps but also by changing the chemical inventory of disks, which extends the conventional understanding that the initial chemistry of natal disks is directly inherited by gas giants, as previously suggested by \citet{OebergEtal2011}.

\subsection{Chemical rings in {MWC\,480}}\label{sec:MWC480}
Among the five targets of MAPS, MWC\,480 stands out as a particularly interesting object for our research. In the continuum, MWC\,480 features an annular gap, D76, at 76 au, potentially indicative of a planet \citep{LiuEtal2019y}. What makes the D76 gap special is that it is spatially coincident with almost all detected emission lines (14 out of the 18 species investigated in \citet{LawEtal2021a}, see \fg{MWC480_obs}). \citet{JiangEtal2022} have found no statistically significant spatial correlations between line emission and continuum in other sources, except MWC\,480. Intriguingly, such concentration of line-emission rings in a dust gap (cavity) configuration is also found in PDS\,70 \citep{FacchiniEtal2021}. 

For these reasons, we have tailored the simulation parameters based on the observational constraints of MWC\,480. Our findings can therefore be directly compared to the observations. In \fg{MWC480_obs}(c), we show the radial profiles of the ALMA continuum emission \citep{SierraEtal2021} and the line emission map of C$_2$H 3-2 \citep{LawEtal2021a}. By applying our model to MWC\,480, the normalized C$_2$H column density profile at $t = 176$\,orbits in the fiducial model is shown in orange for comparison. We convolve the profile with a 0.15\,arcsec beam, which is used in the MAPS data reduction process. Our results are therefore consistent with the observational data. It indicates that an accreting planet can be responsible for the formation of the observed C$_2$H ring, in the D76 continuum gap of MWC\,480 and possibly other molecular rings. 
Further investigation to solidify these findings, e.g., by including a chemical network in our simulations, is worthwhile to confirm these preliminary results.

\subsection{{$\rm C_2H$} rings in other systems}\label{sec:osys}
Apart from MWC\,480, ring-like C$_2$H line-emission substructures are identified and studied in a number of other sources \citep{BerginEtal2016,FacchiniEtal2021,MiotelloEtal2019,BergnerEtal2021,LawEtal2021a}. In these, dust rings and gaps are spatially resolved in the ALMA continuum as well. This raises an interesting question: can the model proposed in this paper also explain other systems?

Within the MAPS samples, HD\,163296 is the only other system where both C$_2$H 3-2 and 2-1 lines present a sharp local maximum inside the D49 continuum gap. 
Besides, a C$_2$H 3-2 ring peaks inside the D117 continuum gap of IM\,Lup, where \citet{PinteEtal2020} suggest a planet candidate depending on its localized kinematics perturbation (kink).
And C$_2$H 1-0 emission radially peaks inside the D100 continuum gap of AS\,209, where a planet candidate is hinted by its kinematic signals \citet{FedeleEtal2023} 
Therefore, our model is possibly applicable to these systems as well.

By statistically studying the existence of C$_2$H emission in disks and the radial profile of their continuum counterpart in 26 disks, \citet{vanderMarelEtal2021c} has suggested a possible correlation between the presence of dust rings outside the CO snowline and the detection of C$_2$H. 
It should be noted that MWC\,480 is an outlier to the correlation, as there is no pebble ring outside its CO snowline, which is approximately at 100\,au.

Under the scenario proposed in this work, the accreting planet may sculpt a pebble ring outside its orbit, at the outer edge of the gap it opened. This ring is therefore located exterior to the methane snowline. Moreover, it is likely to be situated near or beyond the CO snowline as the scenario proposed in this study works at any location beyond the CH$_4$ snowline.
Therefore, our model, which examines the effect of an accreting planet outside the methane snowline on the chemistry of the disk, could potentially provide an explanation for and further complete the tentative correlation identified by \citet{vanderMarelEtal2021c} between the presence of dust rings outside the CO snowline and the detection of C$_2$H.

The sample size of disks with detected C$_2$H remains small, and the available spatial resolution for testing the correlation is limited, which is only possible in MAPS samples. 
Future observations, such as the ongoing ALMA Disk-Exoplanet C/Onnection (DECO) large program, will shed more light on this issue by assisting sample selection. For instance, in the low spatial resolution date of \citet{MiotelloEtal2019}, two sources Sz\,71 (GW\,Lup) and Sz\,129, exhibit tentative features in their C$_2$H 3-2 emission. In both cases, the emission plausibly radially peaks at the D74 and D64 continuum gaps respectively. Interestingly, localized kinematic perturbations consistent with the presence of planet candidates have been detected within these gaps \citep{PinteEtal2020}. Deeper observation with a higher spatial resolution is required to further confirm the link and will be useful for further study.

\subsection{Detectablity of the hypothesized planet}\label{sec:detect}
Observed rings and gaps in protoplanetary disks are often interpreted as evidence of planet formation \citep[e.g.,][]{PerezEtal2019,TociEtal2020a}. Correspondingly, a growing number of embedded planets are being claimed within these disks as well \citep[e.g.,][]{HammondEtal2023,PinteEtal2023}. Therefore, it is important to ask the question where the planet is located? A compelling feature of the planet proposed in our model is that, in order to heat up the surrounding area, it is currently or very recently in the process of accreting.

For instance, MWC\,480 provides an apt illustration of this challenge. Although the Strategic Exploration of Exoplanets and Disks with Subaru (SEEDS) survey has succeeded in directly imaging this object \citep{UyamaEtal2017}, its relatively bright Herbig Ae/Be star status -- with an H band magnitude of 6.26 -- means that detection is still an arduous task. Moreover, identifying any planets in close proximity to the dust ring around MWC\,480 is made all the more difficult due to the possibility of confusion with disk features that may be influenced by advanced image analysis techniques \citep{RameauEtal2017,HaffertEtal2019}.

Detecting planetary accretion signals provides another direct evidence of planets. If a planet accretes gas from the disk, H$\alpha$ line emission originates from the accretion front at the planetary surface. According to the ESO archive, VLT/MUSE \citep{BaconEtal2010} observations have been carried out for more than 50 disks in the past 5 years to search for H$\alpha$ emission \citep[e.g.,][]{XieEtal2020}; yet, so far, PDS\,70 is the only one in which accreting planets have been identified in MUSE \citep{HaffertEtal2019}. A simple explanation, perhaps, is that many planets in disks are accreting too slowly. Therefore, they do not produce detectable H$\alpha$ line emission \citep{MarleauEtal2022}. We might have been fortunate with PDS\,70 b,c, in which the two planets are just above the detection limit. Alternatively, the formation of rings and gaps structure can be independent of the existence of planets \citep[e.g,][]{BaiStone2014,FlockEtal2015,OkuzumiEtal2016,OwenKollmeier2019,OhashiEtal2021,JiangOrmel2021,KuznetsovaEtal2022,TominagaEtal2022}, making the problems more complex.

As discussed in Section \Se{gas_acc}, our rough calculation may face challenges in distinguishing different accretion rate histories above $10^{-5}\,M_{\rm J}\rm\,yr^{-1}$. Additionally, a series of caveats need to be acknowledged, as outlined in \se{limitations}. Despite these limitations, making an educated guess for further investigation is worthwhile. By selecting a mass of $2.3\,M_{\rm J}$ and the corresponding accretion rate of $3\times10^{-5}\,M_{\rm J}\rm\,yr^{-1}$ in our fiducial model, we estimate an H$\alpha$ flux from the planet of $f_{\rm H\alpha} = 1.2\times 10^{-15}\,\rm erg\,s^{-1}\,cm^{-2}$. This estimate is based on the relationship between H$\alpha$ emission and accretion rates presented in Figure 10 of \citet{MarleauEtal2022} and is reasonably above the MUSE detection limit. Our models, therefore, proposed a novel testable approach for searching for planets in disks through their chemistry signals. A searching effort for the potential accreting planets in the future might be also been intriguing in other instruments like MagAO-X \citep{MalesEtal2018} SPHERE/ZIMPOL \citep{CugnoEtal2019}, SCExAO/VAMPIRES \citep{UyamaEtal2020}, and HST \citep{ZhouEtal2021}.

%-----------------------------------------------------------------

\section{Conclusions}\label{sec:conclusions}
In this work, with the newly-developed phase change module \citep{WangEtal2023} of the \texttt{Athena++} code, we perform global multi-fluid hydrodynamic simulations accounting for the sublimation process of the volatile components of (pebble-)accreting planets in a typical protoplanetary disk. The gas accretion of the planet is parameterized with the \citet{MachidaEtal2010a} formula. Our simulations demonstrate how an accreting planet locally heats its surroundings, creates a gap in the disk, and establishes conditions conducive to C-photochemistry. Starting with a $10\,M_\oplus$ core, the planet grows into a Jovian mass planet with accretion rate ${\sim}10^{-5}\,M_{\rm J}\rm\,yr^{-1}$ in our fiducial setup. A sufficiently deep gas gap with a gap depth ${\sim}$30\,\% of the unperturbed disk is opened at the moment when the planet reaches $2.3\,M_{\rm J}$ (\fg{output_default_rad}). Such a bright planet may locally heat its vicinity, elevating the temperature at its Hill radius to ${\sim}45$\,K. A sketch of our model is shown in \fg{chem_sketch}. Our main findings are:

\begin{enumerate}
    \item The enhanced temperature sublimates the major C-carrier ice, methane, surrounding the planet but is not high enough for water and carbon dioxide ice to sublimate. The released methane locally raises the carbon-to-oxygen ratio inside the gap from unity (CO-dominant gas) to ${\sim}2$. The opened gap allows more UV penetration towards the planet's neighborhood. Both high C/O ratio and high UV radiation assist with C-bearing radical (C$_2$H) formation and subsequent assembly of organic compounds.
    \item As a result, a C$_2$H ring is expected to emerge at the dust/gas gap location. The results offer an explanation for the emission of organic molecular lines observed by ALMA in the MAPS large program, where the majority of detected line emission (including C$_2$H, see \fg{MWC480_obs}) peaks inside the D76 continuum gap of MWC\,480. This model may also apply to the C$_2$H ring peak at the D49 continuum gap of HD\,163296.
    \item The outlined model operates when an accreting planet is located beyond the disk CH$_4$ snowline. It predicts the formation of a C$_2$H ring at the planet's location and a pebble ring exterior to it. As the planet may be located beyond the CO snowline as well, the model can potentially account for the correlation between the presence of dust rings outside the CO snowline and the detection of C$_2$H, as observed by \citet{vanderMarelEtal2021c}, see \Se{osys}.
    \item This work demonstrates that the accretion of a giant planet can significantly impact its surrounding astrochemistry environment by locally altering the chemical composition of disks. Due to the different sublimation temperatures, the hot accreting planets can release some volatiles (e.g. methane) while incorporating others (e.g. water), leading to different chemical impacts depending on where the planet is located. This study extends the conventional understanding that the initial chemistry of disks is simply inherited by gas giants, by postulating that planets actively shape their chemical environment of formation.
    \item The finding that a high C/O ratio is the consequence of a gap-opening planet can be inverted to give a rough constraint on the accreting luminosity of the planet. We found that an accretion rate above $10^{-5}\,M_{\rm J}\rm\,yr^{-1}$ is necessary to ensure that the planet is hot enough to release methane, elevate the C/O ratio, and create a prominent C$_2$H ring.
    \item For reference, in the fiducial model, the planet mass reaches $2.3\,M_{\rm J}$ mass when the accreting rate is $3\times10^{-5}\,M_{\rm J}\rm\,yr^{-1}$. The H$\alpha$ flux from the planet is around $f_{\rm H\alpha} = 1.2\times 10^{-15}\,\rm erg\,s^{-1}\, cm^{-2}$, which is potentially detectable via instruments such as VLT/MUSE, SCExAO/VAMPIRES, SPHERE/ZIMPOL, and MagAO-X.
\end{enumerate}

The link between accreting planets and disk astrochemistry proposed by this work is testable with current optical (see \se{detect}), IR (JWST), and (sub)mm (ALMA) wavelength facilities. Conducting planet-hunting campaigns in disks with pronounced chemical rings and continuum gap correlation might therefore be a new avenue to hunt and confirm (proto)planets. At the same time, further investigation into both the spatially resolved disk chemistry and characterization of the atmosphere of the protoplanet in confirmed planet-forming \citep[e.g., for PDS\,70,][]{FacchiniEtal2021,CridlandEtal2023} is of importance to understand how the composition of the planet is built upon, how it influences the disk chemistry in reality, and possibly revealing the formation history of the planet.

Simultaneously, we plan to extend the model to 3D and apply it towards more sources. We also aim to couple our hydro- and thermo-dynamical code with a (reduced) on-the-fly chemical network to obtain more quantitative constraints on chemical interactions between the planet and disk. Further investigation will focus on how the accretion of the planet shapes their atmospheric chemistry, how the effect of planet migration influences the chemical signatures, as well as the impact of multiple planets. Overall, our proposed model opens up exciting new avenues for studying the co-evolution of the disks and protoplanets in planet-forming disks.

\begin{acknowledgements}
    We thank the anonymous referee for their thoughtful and constructive report that improved the initial manuscript.
    H.J. and Y.W. would like to thank Pinghui Huang for fruitful discussions about \texttt{Athena++} code. H.J. thanks Bin Ren and Chen Xie for insightful discussions. H.J. highly appreciates valuable comments from Ilse Cleeves, Alex Cridland, Stefano Facchini, Enrique Macías, Anna Miotello, Giovanni Rosotti, and Claudia Toci. C.W.O. acknowledges support by the National Natural Science Foundation of China (grants no. 12250610189 and 12233004). The authors acknowledge the Tsinghua Astrophysics High-Performance Computing platform at Tsinghua University for providing computational and data storage resources that have contributed to the research results reported within this paper. This work has used \texttt{Astropy} \citep{AstropyCollaborationEtal2013}, \texttt{Matplotlib} \citep{Hunter2007}, \texttt{Numpy} \citep{HarrisEtal2020}, \texttt{Scipy} \citep{VirtanenEtal2020} software packages.
\end{acknowledgements}

% WARNING
%-------------------------------------------------------------------
% Please note that we have included the references to the file aa.dem in
% order to compile it, but we ask you to:
%
% - use BibTeX with the regular commands:
\bibliographystyle{aa} % style aa.bst
\bibliography{ads}
%
% - join the .bib files when you upload your source files
%-------------------------------------------------------------
\begin{appendix} %First appendix

\section{Impact of O-carrier volatiles}\label{app:Orich}
The key of our work relies on the local elevation of the C/O ratio triggered by the accretion luminosity of the planet. In our default setup, we only consider methane in the volatile component. However, oxygen is present in pebbles as well in the form of CO$_2$ and H$_2$O ice. Consequentially, one would expect that oxygen-bearing carriers may also be released from the volatile to the gas phase, which would in contrast lower the local C/O ratio. 

Although it is unlikely to happen due to the relatively high sublimation temperature, see \se{snowline}, we conducted a comparison run by assuming that all icy volatiles in our model is CO$_2$ instead while keeping the other setups the same as the fiducial model. The equilibrium pressure \eq{P_eq} of CO$_2$ follows $T_a = 2571$\,K and $P_{\rm eq,0} = 2.57 \times 10^{12}\,\rm g\,cm^{-1}\,s^{-2}$\citep[][]{FraySchmitt2009} in the simulation. The outcome is straightforward. The typical sublimation temperature of the CO$_2$ ice is $\approx$55\,K (the orange line in \fg{M_dotM_L_T}(d) and see \fg{T_sub}), significantly higher than methane. These temperatures can only be reached in the inner regions of the Hill sphere, but never at the Hill radius. Therefore, as $T_\mathrm{Hill}<50\,\mathrm{K}$, CO$_2$ is almost entirely accreted by the protoplanet and cannot be unleashed into the disk. 

Since water ice has even higher sublimation temperatures (\fg{T_sub}) H$_2$O will also stay on the pebbles in ice form, and be lost to the planet before it sublimates.
As a result, we can confidently assert that our key findings remain unchanged regardless of the water content of the pebble, or indeed any other O-carrier with a significantly higher sublimation temperature than methane.

\section{The $\rm C_2H$ column densities formula}\label{app:n_c2h}
In the study by \citet{BosmanEtal2021b}, a gas-grain chemical network was employed to calculate the 2D ($r-z$) chemical makeup of three MAPS disks. However, due to the lack of a vertical dimension in our simulation, it is challenging to directly compare our results with those of \citet{BosmanEtal2021b}, where the C$_2$H abundance is dominant by photo-chemistry in the upper layer of the disk (e.g., see their Appendix C). Nevertheless, we outline the key features of the \citet{BosmanEtal2021b} model outputs and motivate an empirical fit formula that links the C/O ratio to the C$_2$H abundance, \eq{nc2h}.

\citet{BosmanEtal2021b} concluded that the resolved observations of high but localized C$_2$H column densities suggest the need for locally higher C/O ratios at C$_2$H ring location. Additionally, a correlation is observed between the C/O ratio and CO abundance, indicating that a higher CO abundance corresponds to a lower C/O ratio, vice verse.

Here, we provide a description of how the empirical formula \eq{nc2h} is determined:
\begin{equation*}
    \log_{10} n({\rm C_2H}) = F_C -\log_{10} n({\rm H})+2 \log_2 ({\rm C/O})
\end{equation*}

The dependency of $n({\rm C_2H})$ on $n({\rm H})$, and therefore $\log_{10} n({\rm CO}) \equiv \log_{10} n({\rm H})+5$, is motivated by the anti-correlation observed in Figure 1, Figure 2, and Figure 5 of \citet{BosmanEtal2021b}. These figures demonstrate an anti-correlation between CO column densities and C$_2$H column densities. In particular, Figure 5 shows that the CO column density at the gap location in AS~209 ($\sim 10^{18}$ cm$^{-2}$) is lower than in HD~163296 and MWC~480 ($\sim 10^{20}$ cm$^{-2}$). However, in Figure 2, the C$_2$H column density in AS~209 ($\sim 10^{14}$ cm$^{-2}$) is higher than in HD~163296 and MWC~480 ($\sim 10^{12}$ cm$^{-2}$) when $\rm C/O=1$. Thus, a value of $F_C = 37$ is obtained as the sum of these contributions.

The dependency of $n({\rm C_2H})$ on $\log_2 ({\rm C/O})$ is estimated based on Figure 2 of \citet{BosmanEtal2021b}, which shows that as the C/O ratio increases (or decreases) from 1.0 to 2.0 (or 0.47), the C$_2$H column densities roughly rise (or drop) by approximately two orders of magnitude.

Considering the moderate C$_2$H column densities variation observed in Figure 2 when taking different small dust depletion factors, and the fitting values were visually identified, we leave an uncertainty of approximately one order of magnitude in $F_C$.

\section{Thermodynamic properties of different molecules}\label{app:thermo}
The Clausius–Clapeyron equation is derived from the equality of the chemical potentials between a pure gas-phase compound and its solid phase, which occurs when they are in equilibrium. Taking the perfect gas approximation and neglecting the molar volume of solids, the Clausius–Clapeyron equation can be written as \citep{FraySchmitt2009}
\begin{equation}
    \ln(P_{\rm eq}) = \ln(P_0) + \int_{T_0}^T \frac{\Delta H_{\rm eq}(T)}{RT^2}{\rm d}T
\end{equation}
where $P_0$ is the equilibrium (sublimation) pressure at temperature $T_0$ and
\begin{equation}
    \frac{{\rm d}\Delta H_{\rm eq}(T)}{{\rm d}T} = \Delta C_p(T) = \Delta C_{p,\rm gas}(T) - \Delta C_{p,\rm ice}(T)
\end{equation}
is the equilibrium (sublimation) enthalpy which depends on the heat capacities of gas $C_{p,\rm gas}$ and ice $C_{p,\rm ice}$. If the equilibrium enthalpy is supposed to be independent of temperature, then the equilibrium pressure follows the simple relation
\begin{equation}
    \ln(P_{\rm eq}) = A_0 + A_1/T
\end{equation}
which can be linked to \eq{P_eq} by $P_{\rm eq,0} = e^{A_0}$ and $T_a = -A_1$. We, therefore, take the value of $A_0$ and $A_1$ from Table 5 of \citet{FraySchmitt2009} to calculate the $P_{\rm eq,0}$ and $T_a$ used in our simulation. The original empirical extrapolations formula in \citet{FraySchmitt2009} follows $\ln(P_{\rm eq}) = \sum_{i=0}^{n} A_i T^{-i}$, yet we only take the first two terms to keep the formula still physically meaningful.

\end{appendix}
\end{document}